\documentclass[preprint,11pt,tighten]{aastex}
\usepackage{natbib}
\newcommand{\civ}{C\,{\sc iv}}

\newcommand{\ew}{$W_{\lambda}$}

\newcommand{\hii}{[H\,{\sc ii}]}

\newcommand{\ho}{$H_0$}

\newcommand{\kms}{km\,s$^{-1}$}
\newcommand{\kmsm}{km\,s$^{-1}$\,Mpc$^{-1}$}

\newcommand{\mgii}{Mg\,{\sc ii}}

\newcommand{\oii}{[O\,{\sc ii}]}
\newcommand{\oiii}{O\,{\sc iii}}
\newcommand{\OIII}{[O\,{\sc iii}]}

\newcommand{\qo}{$q_0$}

\newcommand{\aliii}{Al\,{\sc iii}}
\newcommand{\AlIII}{Al\,{\sc iii}\,$\lambda$1858}

\newcommand{\CII}{C\,{\sc ii}\,$\lambda$1335}
\newcommand{\ciii}{C\,{\sc iii}]}
\newcommand{\CIII}{C\,{\sc iii}]\,$\lambda$1909}
\newcommand{\CIV}{C\,{\sc iv}\,$\lambda$1549}
\newcommand{\feii}{Fe\,{\sc ii}}
\newcommand{\feiii}{Fe\,{\sc iii}}
\newcommand{\hei}{He\,{\sc i}}

\newcommand{\HeIIsf}{He\,{\sc ii}\,$\lambda$1640}
\newcommand{\HeIIfs}{He\,{\sc ii}\,$\lambda$4686}
\newcommand{\lya}{Ly$\alpha$}
\newcommand{\LyA}{Ly$\alpha$\,$\lambda$1216}
\newcommand{\lyb}{Ly$\beta$}

\newcommand{\mgiii}{Mg\,{\sc iii}}
\newcommand{\MgII}{Mg\,{\sc ii}\,$\lambda$2798}
\newcommand{\nv}{N\,{\sc v}}
\newcommand{\NV}{N\,{\sc v}\,$\lambda$1240}
\newcommand{\neiii}{[Ne\,{\sc iii}]}
\newcommand{\NeIIIa}{[Ne\,{\sc iii}]\,$\lambda$3869}

\newcommand{\nev}{[Ne\,{\sc v}]}
\newcommand{\NeVa}{[Ne\,{\sc v}]\,$\lambda$3346}
\newcommand{\NeVb}{[Ne\,{\sc v}]\,$\lambda$3426}

\newcommand{\OI}{O\,{\sc i}\,$\lambda$1302}
\newcommand{\OII}{[O\,{\sc ii}]\,$\lambda$3727}
\newcommand{\ovi}{O\,{\sc vi}}

\newcommand{\siiii}{Si\,{\sc iii}]}
\newcommand{\SiIII}{Si\,{\sc iii}]\,$\lambda$1892}
\newcommand{\sio}{Si\,{\sc iv}/O\,{\sc iv}]}
\newcommand{\SiO}{Si\,{\sc iv}/O\,{\sc iv}]\,$\lambda$1400}

\slugcomment{Accepted to the Astronomical Journal on July 11, 2000}

\begin{document}

\title{Active Galactic Nuclei in the CNOC2 Field Galaxy Redshift Survey}

\author{
Patrick B. Hall,\altaffilmark{1,8}
H. K. C. Yee,\altaffilmark{1,8}
Huan Lin,\altaffilmark{2,8,9}
Simon L. Morris,\altaffilmark{3,8}
David R. Patton,\altaffilmark{1,4,8}
Marcin Sawicki,\altaffilmark{1,5,8}
Charles W. Shepherd,\altaffilmark{1}
Gregory D. Wirth,\altaffilmark{4,6,8}
R. G. Carlberg\altaffilmark{1,8}
and Richard Elston\altaffilmark{7,10}
\altaffiltext{1}{Department of Astronomy, University of Toronto, 60 St. George
Street, Toronto, ON M5S~3H8, Canada; E-mail: hall@astro.utoronto.ca}
\altaffiltext{2}{Steward Observatory, The University of Arizona, Tucson, AZ 85721, USA}
\altaffiltext{3}{Dominion Astrophysical Observatory,
Herzberg Institute of Astrophysics, National Research Council, 
5071 W. Saanich Rd., Victoria, BC V8X 4M6, Canada}
\altaffiltext{4}{Department of Physics and Astronomy, University of Victoria,
PO Box 3055, Victoria, BC V8W 3P6, Canada}
\altaffiltext{5}{California Institute of Technology, Mail Stop 320-47, 
Pasadena, CA 91125, USA}
\altaffiltext{6}{W. M. Keck Observatory, Kamuela, HI 96743, USA}
\altaffiltext{7}{Department of Astronomy, The University of Florida,
P.O. Box 112055, Gainesville, FL 32611, USA}
\altaffiltext{8}{Visiting Astronomer, Canada-France-Hawaii Telescope,
operated by the National Research Council of Canada, the Centre National
de la Recherche Scientifique de France and the University of Hawaii.}
\altaffiltext{9}{Hubble Fellow}
\altaffiltext{10}{Visiting Astronomer, Cerro Tololo Interamerican Observatory,
National Optical Astronomy Observatories, operated by AURA Inc., 
under contract with the National Science Foundation.}
}

\begin{abstract}

\small 

We present a sample of 47 
confirmed and 14 
candidate Active Galactic Nuclei (AGN) discovered in the Canadian Network
for Observational Cosmology field galaxy redshift survey (CNOC2).  
The sample consists of 
38 objects identified from broad emission lines,
8 from narrow \nev\ emission,
and 15 candidates from \feii\ or \mgii\ absorption lines,
one of which has been confirmed as a broad-line AGN via infrared spectroscopy.
Redshifts of these AGN range from $z=0.27$ to $z=4.67$, and
the average absolute magnitude is $M_B \simeq -22.25$,
below the quasar/Seyfert division at $M_B = -23$.
Only two of the AGN are detected at radio wavelengths.
We find that only 0.3$\pm$0.1\% of galaxies brighter than $\sim$$M^*$+1
at $0.281 < z < 0.685$ contain broad-line or \nev\ AGN.
We find a total surface density of 270 to 400 AGN~deg$^{-2}$ to $R$=22.09,
comparable to previously published estimates.
About 20\% of these AGN are classified as resolved or probably resolved in CFHT
seeing and might be missed in surveys which target unresolved objects only.
The sample includes several unusual objects:
one with a very strong double-peaked \mgii\ emission line, 
several with unusual emission line properties,
one with an \oiii\,$\lambda$3133 broad absorption line,
and at least one with an optical absorption-line spectrum but broad H$\alpha$
emission in the near-IR.

No color selection criteria were involved in selecting this 
spectroscopically discovered sample.
The sample is also unbiased against objects with luminous host galaxies,
since the spectroscopy preferentially targeted extended objects.
Simple color-color diagram selection criteria can recover $\sim$81$\pm$6\%
of the CNOC2 AGN, but several of the most unusual objects would be
missing from such a color-selected sample.  

In the subsample of broad emission line selected AGN, the average equivalent
widths for \mgii\ and \ciii\ agree with the predictions of previous studies
of the Baldwin effect.  However, the average equivalent widths for \civ\ and
\lya\ are smaller than predicted by previous studies of the Baldwin effect at
lower redshift.  This may imply that the slopes of the \civ\ and \lya\ Baldwin
effects evolve with redshift, steepening with cosmic time.

The broad emission line subsample also shows
a higher incidence of associated \MgII\ absorption than in most previous surveys
and an incidence of associated \CIV\ absorption which may be more similar
to that of radio-selected quasar samples than optically-selected ones.
This may arise from strong absorption being anti-correlated with optical
luminosity or becoming less frequent with cosmic time,
or possibly because our selection method is not biased against objects
with resolved spatial structure
or reddened by dust associated with the absorbing gas.

\end{abstract}

\small


\section{Introduction}	\label{Intro}

Distant Active Galactic Nuclei (AGN), including quasars and Seyferts, have most
often been selected at optical wavelengths as outliers from the stellar locus
defined by broadband color measurements \citep[e.g.,][]{who91}.
The two largest ongoing quasar surveys both
rely on color selection to provide candidate lists \citep{fan99a,boy99}.
Exploration of areas of color space outside reasonable quasar selection
boundaries shows that such color selection misses few quasars 
in magnitude-limited samples \citep{hal96dms1,hal96dms2,hal96dms2err}.  
This allows the incompleteness of
color-selected quasar samples to be calculated and used to determine the
true luminosity function and space density of quasars \citep{who94}.
However, \citet{web95} have presented a radio-selected quasar sample which
includes some very red objects which would not be detectable in traditional
optical surveys in numbers representative of their true space densities.
Also, radio selection of quasars has proven more sensitive than optical
selection to the unusual subclass of quasars known as
Low-ionization Broad Absorption Line Quasars \citep[LoBALs;][]{bec97,bea00}.
Many of these objects show strong absorption from metastable excited states
of \feii\ and \feiii\ (Iron LoBALs) and at least some of them are highly
reddened \citep{ega96,hal97bal}.  Depending on how common substantial dust
obscuration is in quasars \citep{fww99,ke99} and in intervening galaxies
\citep{fp93}, substantial proportions of the true quasar population and of the
populations of some quasar subclasses could be missing from existing surveys.

Therefore it is of interest to obtain samples of AGN selected at different
wavelengths, with a variety of methods with different selection biases
such as morphological selection at high redshift \citep{sar99a,bwa99} 
and variability selection \citep{czm93,hv95},
and especially with as few selection criteria as possible.
One obvious way to achieve this latter goal in the optical is through purely
magnitude-limited samples.  The low surface density of AGN relative to
stars and galaxies means that only very small samples of AGN with minimal
selection criteria have been assembled in this manner to date.  
\citet{sch91} and \citet{col91} discuss samples of five and two AGN,
respectively, discovered using very low resolution spectroscopy
($\sim$50-75\,\AA\ FWHM).
\citet{sch96a}, \citet{coh99a} and \citet{coh00} discuss samples of six, three
and three AGN, respectively, discovered in moderate resolution spectroscopic
redshift surveys.
All these AGN were selected only by magnitude and broad-line detectability.
The emission line equivalent widths of the \citet{sch96a} sample
do not differ significantly from those of \citet{hs90} or \citet{fra91}.
However, five of the six quasars have continuum spectral indices redder
than the median of the large sample of \citet{fra91}, although this difference
is not statistically significant given the small sample size.

In this paper we present a sample of AGN discovered spectroscopically in the
Canadian Network for Observation Cosmology field galaxy redshift survey
(hereafter CNOC2).
This sample was selected without reference to broadband color and is the
largest AGN sample yet assembled directly from moderate-resolution spectroscopy.
We discuss the CNOC2 dataset in \S\ref{Data},
present the AGN sample in \S\ref{Selection},
provide notes on individual AGN in \S4--6,
discuss the properties of the sample in \S\ref{Discussion},
and summarize our conclusions in \S\ref{Conclusions}.
We adopt \ho=50~\kmsm, \qo=0.5\ and $\Lambda$=0 for ease of comparison to
previous AGN studies.

\section{Data}  \label{Data}

The CNOC2 dataset is discussed in \citet{yea00} 
and only relevant details are summarized here \citep[cf.][]{cnoc2.1}.
$UBgR_cI_c$ imaging and spectroscopy were obtained using the
Multi-Object Spectrograph (MOS) on the Canada-France-Hawaii Telescope (CFHT),
with the $g$ photometry calibrated to $V$ magnitudes.
The survey covers four {\em patches} on the sky, each of which is a mosaic of
17 to 19 contiguous MOS {\em fields} of size 
7.3\arcmin\ or 8.3\arcmin\ $\times$ 9\arcmin.
Spectroscopy was obtained for $\sim50$\% of $R<21.5$ galaxies
using two slit masks per field with two exposures 
per mask (20 to 40 minutes each), slits of width 1\farcs3 and minimum length
10\farcs5, and $\sim$100 objects per mask.  
The final rebinned 1-D spectra cover 4390 to 6292.21~\AA\ at 4.89\,\AA/pixel
and resolution $\sim$14.8\,\AA.
The slit masks were created using images obtained and catalogued using 
PPP \citep{yee91} during each run, and slits were assigned only to objects 
morphologically classified by PPP as galaxies or probable galaxies 
in the $R$ images.  
However, there were a few cases where morphologically classified stars
were observed spectroscopically:
if they serendipitously fell on the slits,
if they were located in a region of the mask empty of target galaxies
and a slit could be added without affecting previously assigned slits,
or if they were accidentally classified as galaxies in the ``summit'' catalogs
used to create the masks, but not on the final catalogs constructed after
exhaustive quality checks.
This misclassification could happen either because they were
confused with other very nearby objects or because they were located in
poorly focused areas where careful PSF construction is necessary for 
accurate star-galaxy separation.

Galaxy redshifts were determined by cross-correlation with a set of 
three templates: one absorption-line, 
	one emission + absorption, 
	and one emission-dominated.  
The choice of template to use,
and thus the spectral class and redshift assigned to the galaxy,
was only made after an interactive review which involved examining
the direct and spectroscopic images as well as the final 1-D spectra.
This process enabled broad lines or other unusual spectral features to be 
flagged.  

In the CNOC2 catalog (2000 April version), 
a total of 6178 galaxies and AGN have measured redshifts,
226 objects were spectroscopically classified as stars, 
and 564 objects remain unidentified despite having a spectrum with a 
signal-to-noise ratio (SNR) nominally sufficient for classification.  
By design, the vast majority of objects with spectra are morphologically
classified galaxies.  Despite the large variation of the PSF across
the MOS field, the morphological star-galaxy separation is quite robust: 
only 39 morphologically classified galaxies 
are spectroscopically classified as stars,
and of the 288 morphologically classified stars for which spectra adequate for
classification were obtained, 146 are spectroscopically classified as stars,
65 as galaxies and 17 as AGN or candidate AGN, while 60 remain unidentified.

\section{Selection of AGN}	\label{Selection}

The CNOC2 data reduction pipeline includes several interactive steps
in which AGN were identified and noted in a single log file:  during spectral
extraction, during cross correlation, and during final redshift verification
\citep{yea00}.
The 2-D and 1-D spectra were examined by at least two different people.
Spectra of all candidate AGN, all spectroscopically unclassified objects,
and all emission-line objects 
were then examined by eye by one of us (PBH) to assemble the final AGN sample.
AGN were identified not only from broad emission lines in their spectra
(see Table~\ref{tab_spec_bel} for the most common lines),
but also from narrow \NeVa\ or \NeVb\ emission (\S\ref{nevsample})
and from \feii\ or \mgii\ absorption lines (\S\ref{abssample}).
The identification of narrow line AGN at $z\lesssim0.6$ 
(e.g., from emission line ratios) 
is beyond the scope of this paper.
Possible causes of spurious lines include zeroth order emission, cosmic rays,
spatially and spectrally adjacent slit overlaps, and flatfield residuals.
These are ruled out in each case by visual inspection of the 2-D spectra.

Redshifts for emission line objects without \OII\ emission or other 
lines present in the cross-correlation templates 
were measured by fitting Gaussians to all 
absorption or emission lines, assuming $1\sigma$ redshift uncertainties of 0.425
times the FWHM of the line, and taking the average redshift where appropriate.
Because of the limited spectral coverage, at certain redshifts
either no strong line or only one is expected to be visible.
At $1.24\lesssim z\lesssim 1.31$ neither \MgII\ nor \CIII\ will be visible.
There is a potential degeneracy between \mgii\ at $0.58<z<0.94$ and \ciii\ at
$1.31<z<1.84$, although it can be broken at $z \leq 0.685$ if \OII\ is present.
The \mgii/\ciii\ degeneracy can also be broken if associated absorption is
present since it cannot appear in \ciii; however, absorption does occasionally
appear in \AlIII\ so a redshift derived from this distinction is not as definite
as a two-line redshift.
It is also possible that a single line could be \CIV\ if other lines that would
be visible in that case (e.g., \CIII, \SiO, \CII, \OI) 
are weak, but from inspection of the spectra we do not believe this could be the
case for any of the CNOC2 AGN except perhaps 0223.021739 (\S\ref{belsample}).

\subsection{Magnitudes, Coordinates, and Literature IDs} \label{magscoordsids}

In Tables~\ref{tab_spec_bel}--\ref{tab_spec_abs}
we provide the redshifts, spectroscopic $R_{cor}$ values
\citep[$R_{cor}$ is the signal-to-noise ratio of the redshift correlation peak, cf.][]{yec96},
morphological classifications (see \S\ref{morphs}), and emission-line
equivalent widths (see \S\ref{EWs})
for all CNOC2 AGN.
All CNOC2 objects are referred to by a four-digit patch code
plus a six-digit field+object code after a decimal point;
e.g., 2148.050488 is object 488 of field 5 of patch 2148 \citep{yea00}.
In Table~\ref{tab_phot} we provide coordinates, apparent {\em UBVRI} magnitudes
corrected for Galactic extinction as outlined in \citet{yea00},
and absolute $B$ magnitudes $M_B$ for all AGN.  
Optical distortions in the MOS instrument prevent direct transformation of 
CCD x and y to RA and Dec coordinates, but a simple model of this distortion
is used to produce reasonably accurate relative coordinates in the CNOC2
photometric catalog \citep{yea00}.
To convert these coordinates to RA and Dec, each CCD field was first matched
to the USNO-A1.0 catalog \citep{usnoa1} to provide rough RA and Dec coordinates,
and then each patch was fit as a whole to the USNO-A1.0 catalog.
The resulting coordinates are accurate to 
$\pm$1\arcsec\ for most objects.
$M_B$ values were computed from the observed $B$ magnitudes
($V$ magnitudes for the two $z>4$ objects)
using the $k$-corrections of \citet{cv90}.
Note that these $k$-corrections will be inappropriate
for any absorption-line objects which are actually star-forming galaxies.

A search was made for objects within 1\arcmin\ of each object's position using
NED,\footnote{The NASA/IPAC Extragalactic Database (NED) is operated by the Jet 
Propulsion Laboratory, California Institute of Technology, under contract with 
the National Aeronautics and Space Administration.}
the NVSS catalog \citep{con98},
and (for the 0223+00 and 0920+37 patches only) the FIRST catalog \citep{bwh95}.
Only two objects are coincident with sources previously known at any wavelength
within the combined positional uncertainties:
both 0223.130495 and 2148.102026 are radio sources.
They are the only objects of ours detected by NVSS or FIRST,
and both are radio-loud.
See \S\ref{belnotes} for discussion of the identifications.

\subsection{Redshift and Absolute Magnitude Distributions}	\label{zMB}

\begin{figure}
\epsscale{2.0}
\plottwo{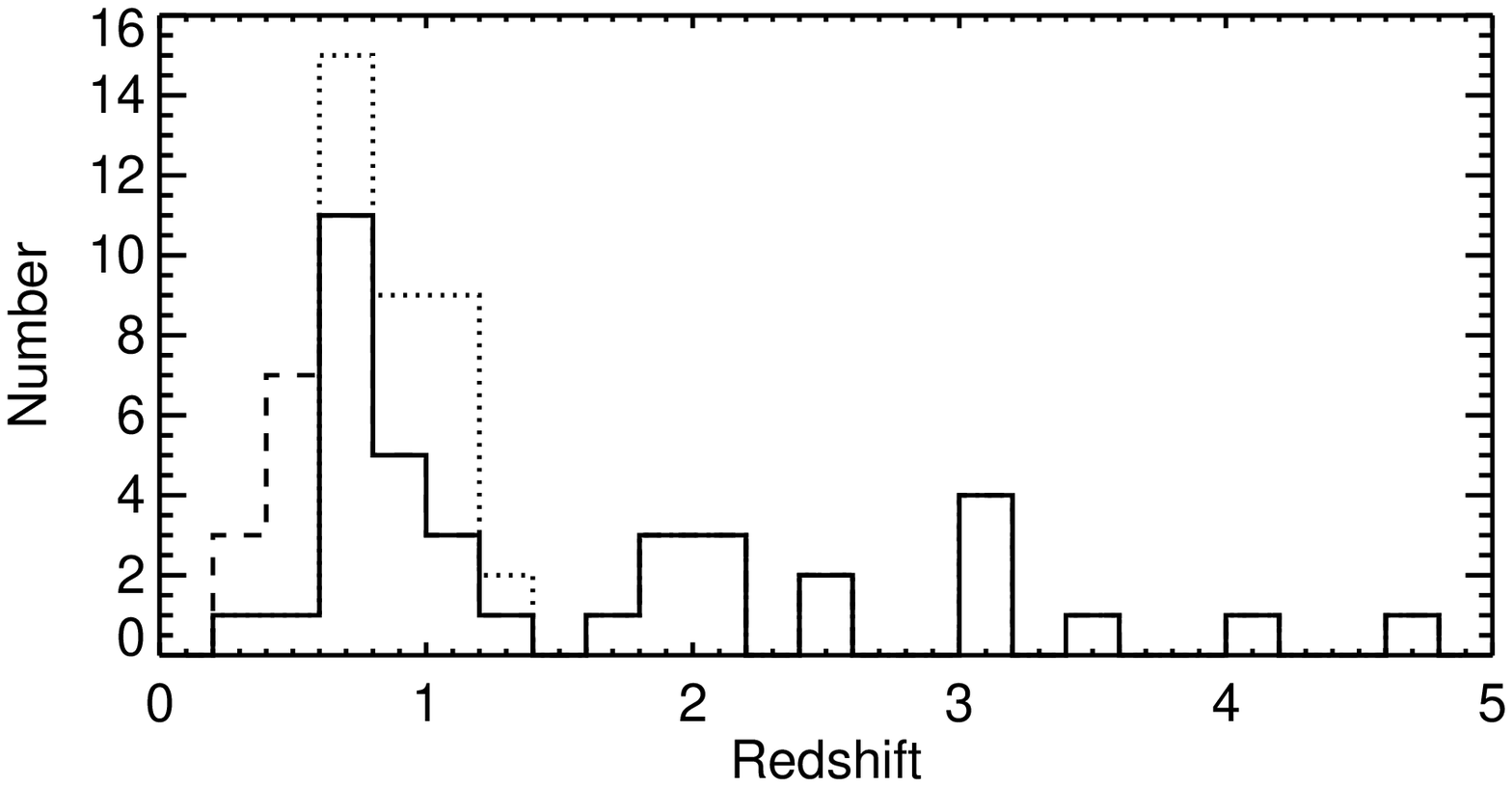}{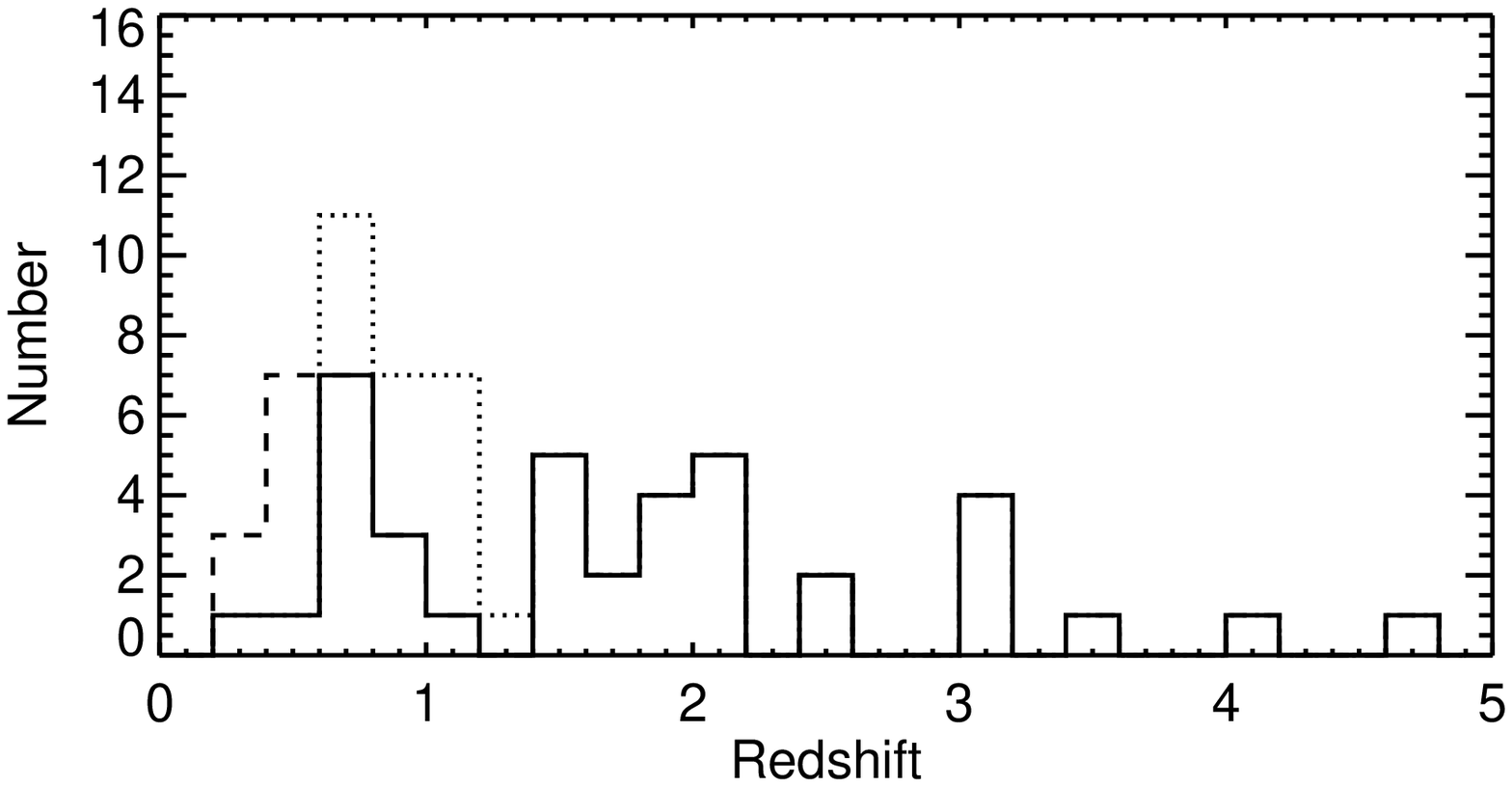} 
\caption[]{
\singlespace
Redshift histogram for CNOC2 AGN.
The solid line shows only broad emission-line (BEL) selected objects;
the dashed line, those plus \nev\ selected objects;
and the dotted line, all AGN, including absorption-line selected candidates.
Figure~\ref{f_zhist}a shows the distribution of our adopted redshifts,
and Figure~\ref{f_zhist}b the distribution when every single-emission-line
object of uncertain redshift is given its \ciii\ redshift instead of \mgii\ 
(or the reverse in two cases). 
}\label{f_zhist}
\end{figure}

Figure~\ref{f_zhist}a shows the redshift histogram for all AGN.
The solid line shows only broad emission-line (BEL) selected objects, 
the dashed line those plus \nev\ selected objects,
and the dotted line all AGN, including absorption-line selected candidates.
We consider all the BEL AGN to be confirmed AGN. 
The only confirmed AGN among the absorption-line objects is 2148.050488
at $z=1.328$ (see \S\ref{irspec}); the rest are merely candidate AGN.
The two highest redshift AGN in the sample are the third and fourth
serendipitously discovered at $z>4$ \citep{mcc88,ssg94}, but considerably
more spectra were obtained to find these objects than in the previous cases.
Figure~\ref{f_zhist}b shows the redshift histogram with every 
single-emission-line object of uncertain redshift given its \ciii\ redshift
instead of \mgii\ (or the reverse in two cases).  Both histograms show the
expected scarcity at $1.24\lesssim z \lesssim 1.31$ where no line is expected
to be visible.  Spectroscopy with wider wavelength coverage will be needed to
pin down the redshifts of the single-emission-line objects.

\begin{figure}
\epsscale{1.0}
\plotone{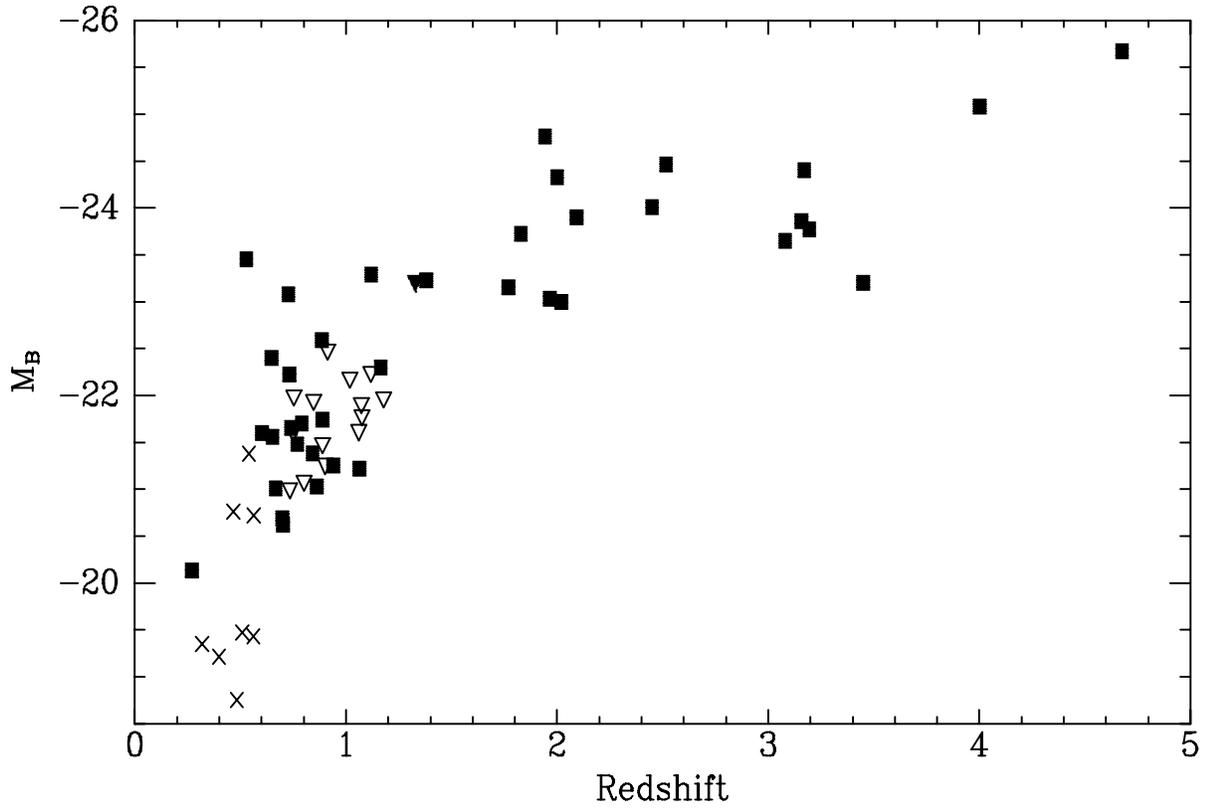} 
\caption[]{
\singlespace
Absolute rest-frame $B$-band magnitudes vs. redshift for the CNOC2 AGN.
Filled squares are broad emission-line (BEL) selected objects,
crosses are \nev\ selected objects, 
filled triangles are confirmed absorption-line selected AGN,
and open triangles are candidate absorption-line selected AGN.
Note that redshifts of some one-line objects suffer from a degeneracy
between \mgii\ and \ciii\ and may be incorrect.
}\label{f_mz}
\end{figure}

Figure~\ref{f_mz} shows absolute $B$ magnitudes vs. redshift for all CNOC2 AGN.
On average $M_B=-22.25$ for confirmed AGN.  
This is below the traditional division between quasars and Seyferts
at $M_B=-23$ (\ho=50, \qo=0.5).
These magnitudes are based on an average $k$-correction (\S\ref{magscoordsids}).
Individual quasars' magnitudes may show variations of up to $\pm$0\fm5 from
this average, in line with the spread seen in their $U-B$ and $B-V$ colors
at low redshift (\S\ref{Colors}).
Host galaxy contamination may also be significant
in some objects, especially for the fainter AGN whose magnitudes may be
comparable to those of their hosts. 

\subsection{Spectra}	\label{spectra}

\begin{figure}
\epsscale{1.5}	
\plotone{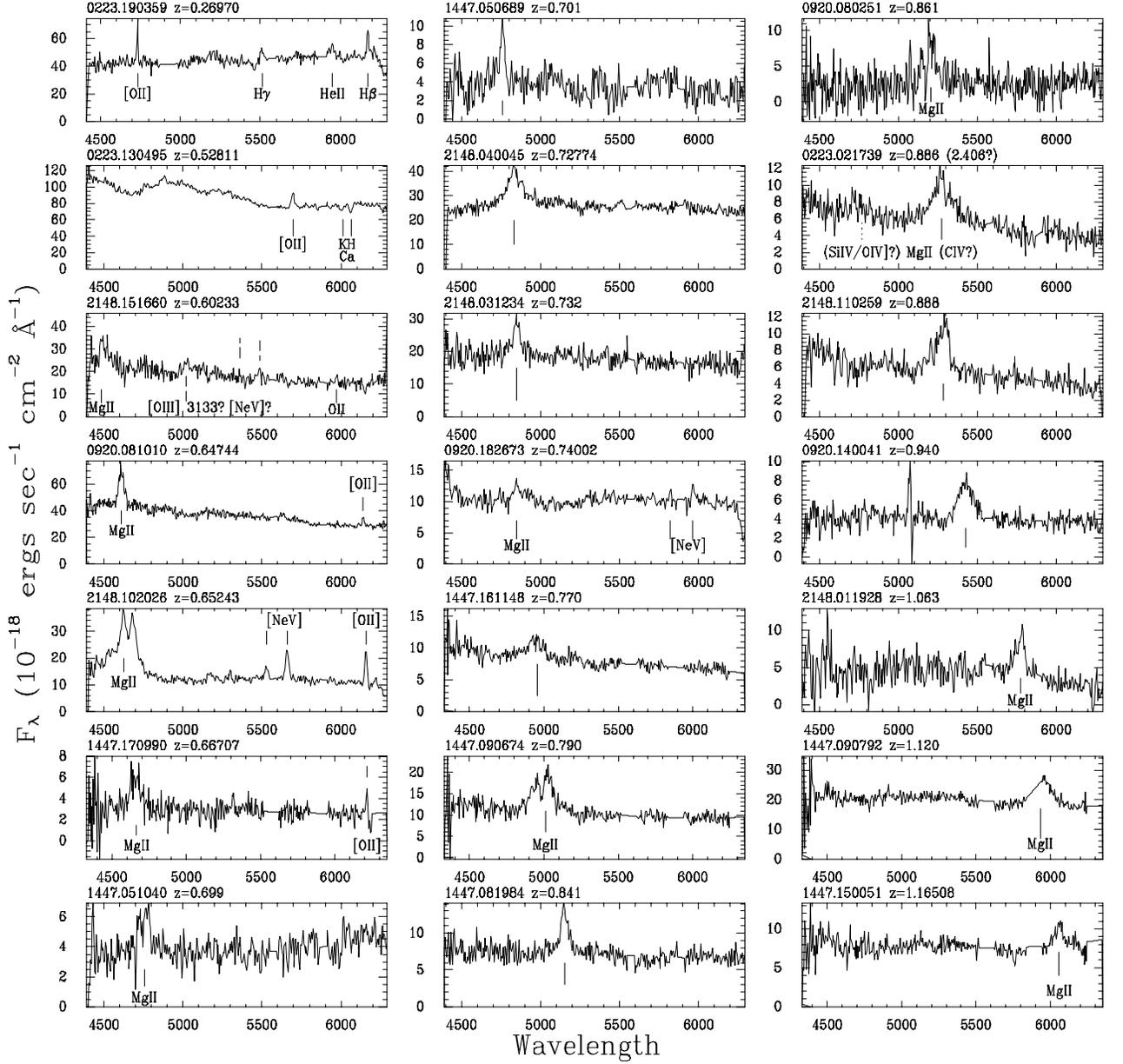} 
\caption[]{
\singlespace
Spectra of the thirty-eight broad emission line selected CNOC2 AGN,
in columns of increasing redshift from top to bottom and left to right.
We consider all these to be confirmed AGN, but many with only one detected line
suffer from a redshift degeneracy between \MgII\ and \CIII.
Detected features are marked with a solid line and the line identification,
possible features are marked with a dashed line, and
dotted lines show the positions of expected but missing or undetected features.
See \S\ref{belnotes} for discussion of individual objects.
}\label{f_spectra_bel1}
\end{figure}

\begin{figure}
\epsscale{1.5}
\plotone{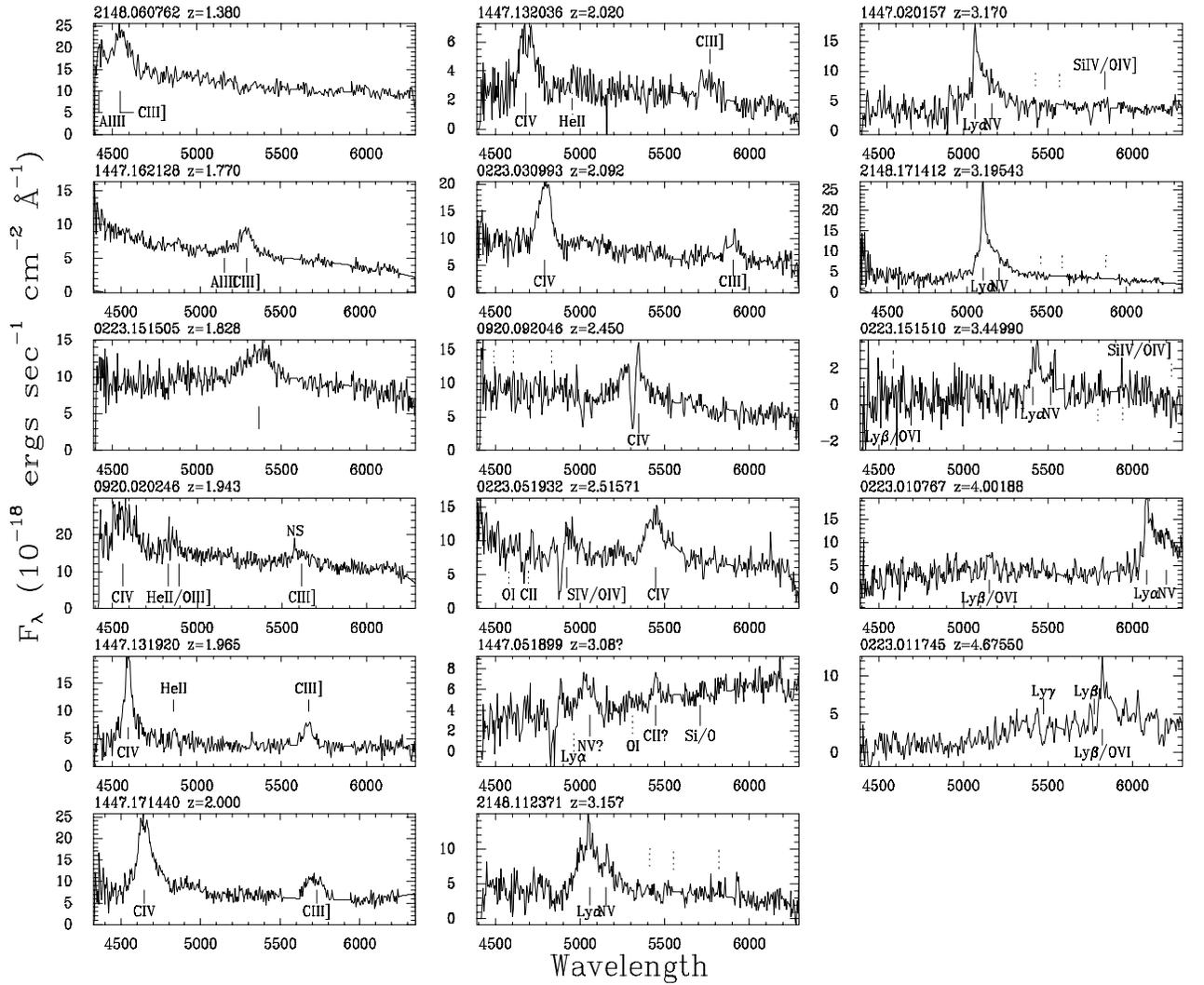} 
\caption[]{
\singlespace
Spectra of the thirty-eight broad emission line selected CNOC2 AGN, continued.
See Fig.~\ref{f_spectra_bel1} for details and
\S\ref{belnotes} for discussion of individual objects.
}\label{f_spectra_bel2}
\end{figure}

\begin{figure}
\epsscale{1.1}
\plotone{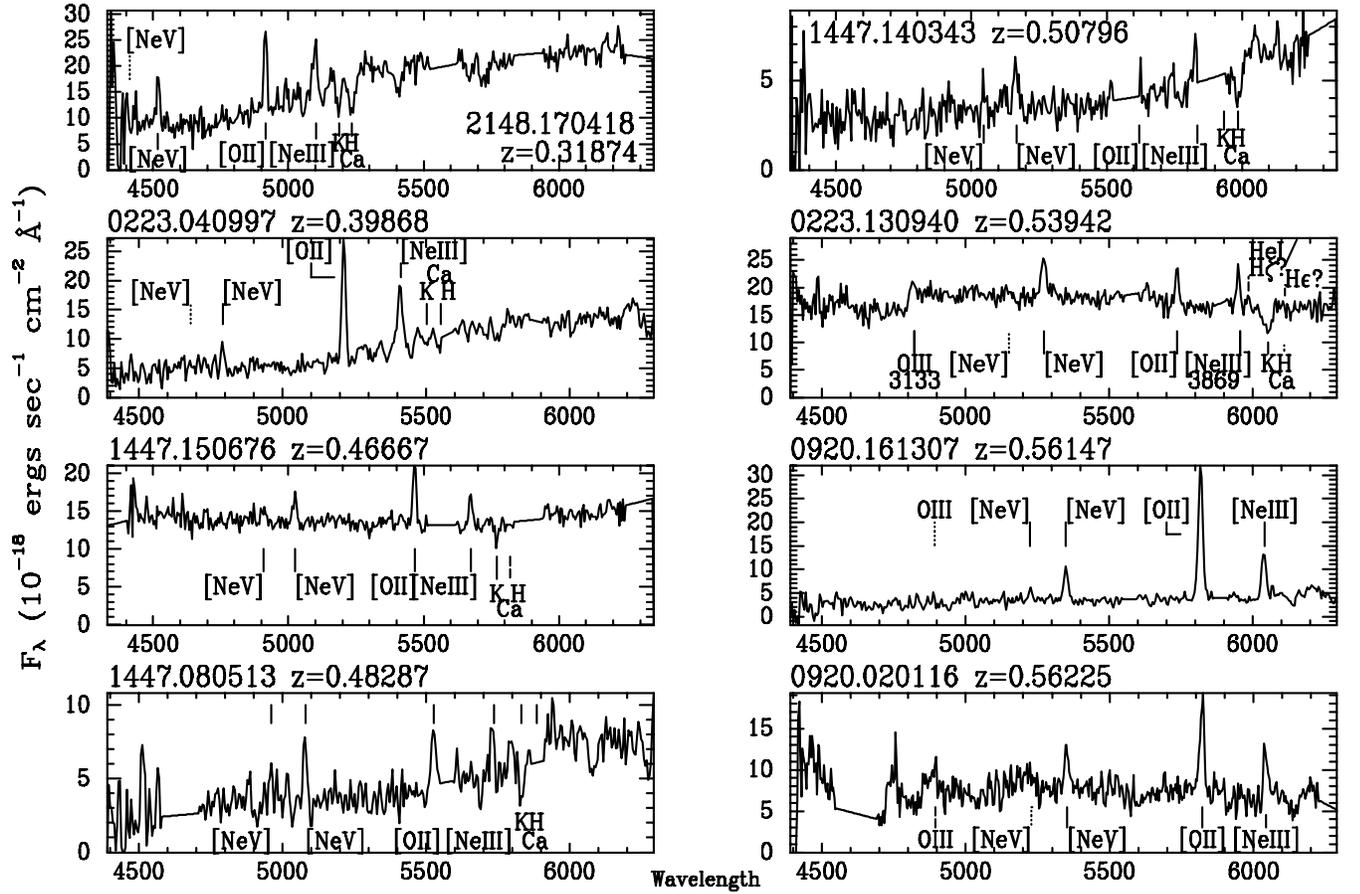} 
\caption[]{
\singlespace
Spectra of the eight CNOC2 AGN selected by their \nev\ emission.
See Fig.~\ref{f_spectra_bel1} for details and
\S\ref{nevnotes} for discussion of individual objects.
}\label{f_spectra_nev}
\end{figure}

\begin{figure}
\epsscale{1.1}
\plotone{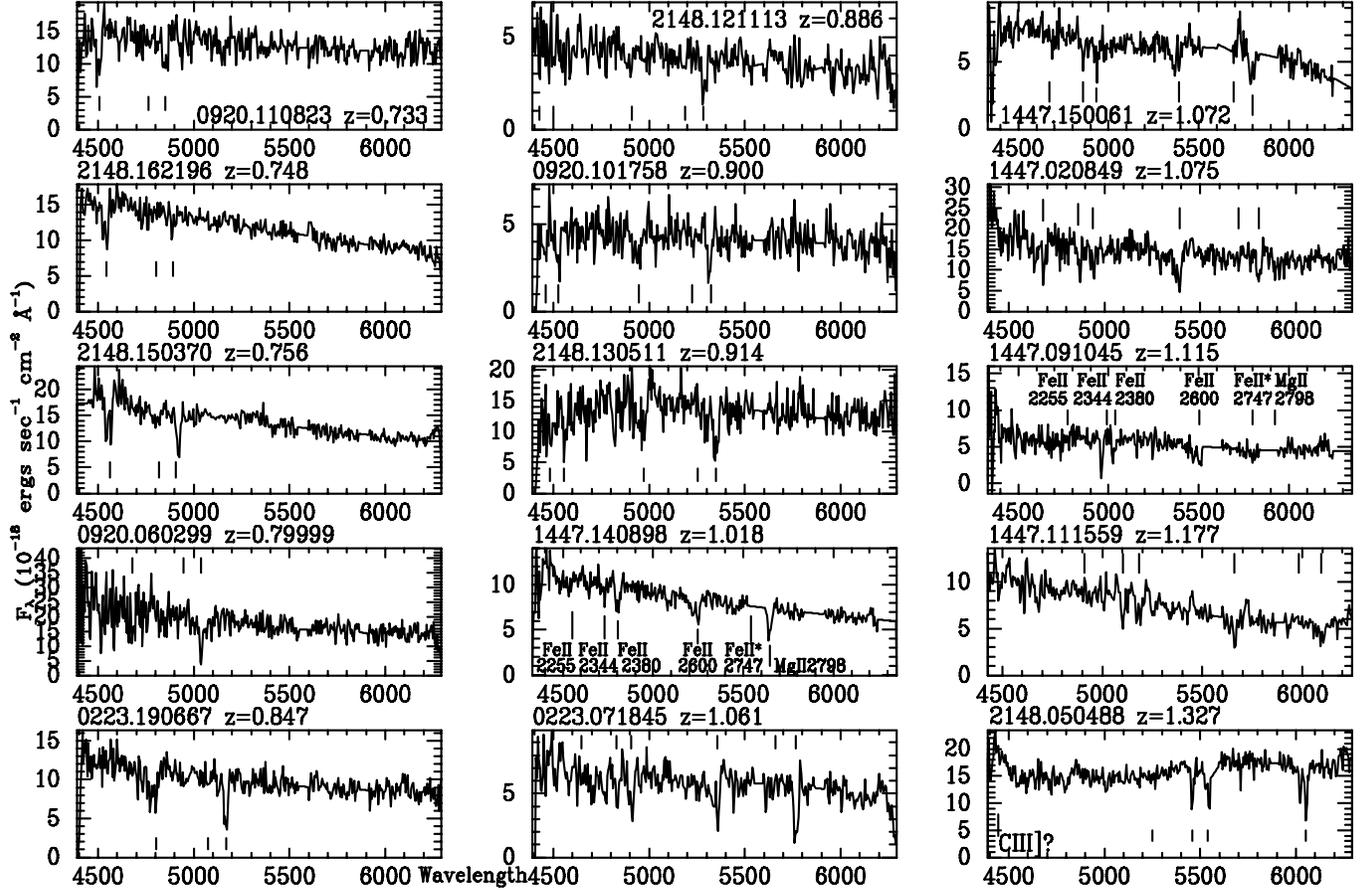} 
\caption[]{
\singlespace
Spectra of the fifteen absorption-line-selected CNOC2 AGN candidates,
in columns of increasing redshift from top to bottom and left to right.
The positions of common lines (or line pairs) of \feii\ and \mgii\ are marked
with vertical ticks in each spectrum (and are labelled in two spectra for reference),
regardless of whether the line is present in that spectrum.
}\label{f_spectra_abs1}
\end{figure}

Figures~\ref{f_spectra_bel1}--\ref{f_spectra_bel2} 
show the spectra of the broad emission line selected AGN, 
Figure~\ref{f_spectra_nev} the spectra of the \nev\ emission selected AGN,
and 
Figure~\ref{f_spectra_abs1} 
the spectra of the absorption line selected AGN candidates.
Objects are displayed in order of increasing redshift.
Where multiple spectra of the same object were available,
they were combined with a simple average.
Notes on individual objects 
are provided in the following sections.

\section{Broad Emission-Line AGN}	\label{belsample}

In the following discussions we use the term {\em neighbor} to mean a galaxy
close to the AGN on the sky, and reserve the term {\em companion} for a galaxy
known to be at the AGN redshift.
When discussing the morphology of the CNOC2 AGN host galaxies, it must be
remembered that the MOS imaging has a coarse pixel scale, typically
0\farcs438/pixel, and that there is significant defocusing across the chip,
resulting in image quality and resolution that is not particularly good in
general.  Using the CFH12k mosaic CCD camera on CFHT, we have obtained higher
quality images of most of the CNOC2 area as part of the Red-Sequence Cluster
Survey \citep{gy00a,gea00}.
Where possible, we discuss the morphology of our AGN based on these images,
since they are at least a magnitude deeper than the CNOC2 imaging,
have better resolution (0\farcs6-0\farcs8 FWHM),
and are better sampled (0\farcs2/pixel).

\subsection{Notes on Individual Broad Emission-Line AGN}	\label{belnotes}

\paragraph{0223.190359} 
This $z=0.27$ object shows narrow \oii\ plus broad \HeIIfs, H$\gamma$, 
and possibly H$\beta$ emission.
Morphologically it is a well resolved, high surface brightness, round galaxy.

\paragraph{0223.130495} 
This $z=0.528$ object shows moderately broad (FWHM$\simeq$1200\,\kms)
\oii\ emission and a very blue continuum with strong broad features.  
\feii\ emission seems the likely identification for these features.  
However, the only obvious correspondences with the \feii\ emission spectrum
of the narrow-line quasar I~Zw~1 \citep{lao97} are 
the break at 4565\,\AA\ observed (2987\,\AA\ rest) 
	with the red edge of the \feii\ UV multiplets 60 and 78,
and the emission at 4880\,\AA\ observed (3193\,\AA\ rest) 
	with emission from \feii\ optical multiplets 6 and 7.
There is some correspondence with the theoretical \feii\ spectra of
\citet[][their Figure~12b]{ver99}, particularly at 3100--3500\,\AA.
Alternatively, the possible emission features blueward of 4565\,\AA, at
4735--4785\,\AA, and at 4845--4900\,\AA\ might be identified with the 2970, 3130,
and 3200 \AA\ features seen in the composite quasar spectrum of \citet{fra91}.
Morphologically, the object has a nucleated center, but is clearly resolved.

0223.130495 is identified with the radio-loud quasar PKS 0222+000,
a 278~mJy 
1.4~GHz source barely resolved by FIRST \citep{bwh95}.
The spectrum of \citet{wil83} is of too low SNR to show the unidentified
broad features seen in our spectrum.  
\citet{cbj76} reported variable colors in this object,
measuring $m_{pg}$=18 compared to $m_{pg}$=19 in \citet{bf85}, 
and \citet{dun89} reported $B$=19.8$\pm$0.5 and $R$=18.3$\pm$0.5 
compared to the $B$=18.93$\pm$0.03 and $R$=18.27$\pm$0.03 that we measure.
This object may have developed the unidentified broad features in the blue
in the $\sim$8 years (rest frame) between the observations of \citet{dun89} 
in 1984 June -- 1985 September and our observations in 1996 October.

\paragraph{2148.151660}  
The redshift of this $z=0.60233$ object comes from possible weak \oii\ emission,
corroborated by emission from \MgII\ and the 2970\,\AA, 3130\,\AA, and
3200\,\AA\ features of \citet{fra91}, which may be \feii\ features 
\citep{lao97} or \oiii\,$\lambda$3133 in the case of the 3130\,\AA\ feature.
There is also possible weak \nev\ emission.
The object is compact but clearly extended with a relatively large, low surface
brightness halo in the direct images.

\paragraph{0920.081010}	
This $z=0.64744$ object shows broad \mgii\ and narrow \OII\ emission.
The redshift was derived from the latter.
Morphologically the object may be slightly resolved on the MOS imaging.
CFH12k $Rz'$ imaging (\S\ref{belsample})
definitely resolves the object, and in addition shows it to be surrounded
by faint diffuse emission which is strongest to the NNW.

\paragraph{2148.102026} 
This $z=0.65243$ object exhibits
both \NeVb\ and \NeVa\ as well as narrow \oii\ and broad \mgii.
The \mgii\ line has an exceptionally large equivalent width.
It either has associated absorption redward of the narrow-line
emission redshift or has a double-peaked profile with the blue peak at the
narrow-line redshift and the red peak $\sim$3700$\pm$100\,\kms\ redward of it.
Possible emission is also seen at $\sim$3130\,\AA\ and $\sim$3200\,\AA,
which could correspond to the unidentified features of \citet{fra91}
or to \oiii\,$\lambda$3133 
in the first case.
The object is in a region of poor focus and large PSF distortions.
Thus its proper classification is difficult to determine, but it does
appear more extended than a nearby star of about the same brightness,
so its classification as stellar may be in error.

2148.102026 is identified with the ``symmetric double'' radio source 
TXS 2149--059 \citep{dou96}, and is also detected by NVSS as a resolved
176.4$\pm$6.1 mJy 1.4~GHz source \citep{con98}.

About 20 examples of double-peaked emission lines in AGN are known,
mostly in broad-line radio galaxies \citep{eh94,eh98}.  
For this reason, and given the good agreement of the blue peak and the 
narrow-line redshift, we believe the double-peaked profile is the more 
likely explanation for this object's \mgii\ emission profile.  
However, double-peaked \mgii\ is rare; for example, only one or two
possible candidates are present among over 500 radio-selected AGN with
\mgii\ coverage from the FIRST Bright Quasar Survey \citep{wea00}.  
Thus this object is clearly worthy of further study.

\paragraph{1447.170990} 
This $z=0.667$ object shows broad \mgii\ and narrow \OII\ emission.
There narrow emission at 5313.9\,\AA\ observed (3187.6\,\AA\ rest)
is probably spurious.
It has a neighbor or extension visible about 1\arcsec\ to the W in the $g$,
$R$ and $I$ images, but is nonetheless unresolved and classified as a star.

\paragraph{1447.051040} 
This object has one broad line which is assumed to be \mgii\ at $z=0.699$
due to the presence of associated absorption.
If the line were \civ\ at $z=3.07$, \ciii\ would have to be abnormally weak.
Morphologically the object appears to be stellar but with faint symmetric fuzz,
and it is classified as a galaxy.
There is a galaxy of unknown redshift only 1\farcs8 to the NW (1447.051044).

\paragraph{1447.050689} 
This object has one moderately broad line which is assumed to be \mgii\ at 
$z=0.701$, although an identification of \ciii\ at $z=1.493$ cannot be ruled out.

\paragraph{2148.040045}  
This object has one broad line which is assumed to be \mgii\ at $z=0.728$,
although an identification of \ciii\ at $z=1.532$ cannot be ruled out.
This object has a bright nucleus but is clearly extended and asymmetric,
with emission to the NW and fainter emission to the NE.

\paragraph{2148.031234}  
This object has one broad line which is assumed to be \mgii\ at $z=0.732$,
although an identification of \ciii\ at $z=1.539$ cannot be ruled out.
This object is clearly extended, and there is a galaxy of unknown redshift
4\farcs1 to the SE (2148.031219).

This object has been serendipitously imaged in an ongoing {\em HST} WFPC2
$F814W$ band survey of CNOC2 galaxy pairs 
\citep{pat01}\footnote{Based on data from the NASA/ESA Hubble
Space Telescope, obtained at the Space Telescope Science Institute, 
operated by the Association of Universities for Research in Astronomy, Inc.
under NASA contract No. NAS5-26555.}.  
It is clearly resolved into a pointlike AGN and a diffuse, apparently bulgeless
host galaxy whose isophotes are not centered on the AGN.
It could be a face-on spiral galaxy or possibly a disturbed system.
The galaxy 4\farcs1 to the SE is resolved, quite compact, and also asymmetric,
but there is no clear evidence for any interaction between the two galaxies.

\paragraph{0920.182673} 
This object is definitely resolved, possibly with faint extended emission
(the unresolved red object about 3\arcsec\ to the SW, 0920.182662,
did not affect its classification).
This object shows \NeVa\ and \NeVb\ emission at $z=0.74$ and \MgII\ blueshifted
by --1350$\pm$150\,\kms, a large but not unprecendented blueshift \citep{tf92}.
Note that we cannot completely rule out a very narrow-lined quasar at $z=2.129$ 
with \CIV, \AlIII, \CIII\ and possibly \SiO\ emission in the extreme blue.
However, the extreme weakness of \civ\ and \ciii\ and \aliii\ in that case,
the low probability of such a large host galaxy at $z\sim2$,
and the better fit of the two reddest lines to the wavelengths of \NeVa\ and
\NeVb\ than \AlIII\ and \CIII\ all lead us to adopt the $z=0.74$ ID.

\paragraph{1447.161148} 
This object has one broad line which is assumed to be \mgii\ at $z=0.77$,
although an identification of \ciii\ at $z=1.59$ cannot be ruled out.
There is a neighboring galaxy of unknown redshift located 2\farcs4 to the NE
(1447.161160).  It may be responsible for this object being classified as a
probable galaxy, or the object may in fact be slightly resolved.

\paragraph{1447.090674} 
This object has one broad line which we identify as \mgii\ at $z=0.79$
since it shows associated absorption; 
it cannot be \civ\ since both \sio\ and \ciii\ would be missing in that case.
Morphologically this is a slightly resolved, compact object of high surface
brightness.

\paragraph{1447.081984} 
This object has one moderately broad line which is assumed to be \mgii\ at 
$z=0.841$, although an identification of \ciii\ at $z=1.698$ cannot be ruled out.
This unresolved object was added to its spectroscopic mask by hand to fill an
empty area.

\paragraph{0920.080251}	
This $z=0.861$ object shows a single broad emission line
	(which falls atop the 5200\,\AA\ night sky line)
which we identify as \mgii\ due to the presence of associated absorption.  It
is likely unresolved and forms a close pair with a faint red galaxy to the SSE.
The objects were not separated by PPP and were classified as a single galaxy.

\paragraph{0223.021739}	 
This object shows one broad emission line with associated absorption.  We
identify it as \mgii\ at $z=0.886$, but \civ\ at $z=2.407$ cannot be ruled out.
There is possible weak \sio\ emission which might corroborate the \civ\
interpretation, but it is slightly blueward of the expected wavelength.
The object's colors are consistent with either redshift.
Morphologically it is unresolved.

\paragraph{2148.110259}  
This object has one broad line which is assumed to be \mgii\ at $z=0.888$,
although an identification of \ciii\ at $z=1.767$ is possible.
It is classified as a probable galaxy, but appears barely resolved at best.
There is a galaxy of unknown redshift 2\farcs5 to the ENE (2148.110273),
but the two were 
separated by PPP so it is unlikely
to be responsible for the AGN's classification as a probable galaxy.

\paragraph{0920.140041} 
This $z=0.94$ object shows only one line, which is almost certainly \mgii.  
\ciii\ at $z=1.84$ cannot be ruled out entirely, since although there is no
sign of the \civ\ emission expected in that case, it would be present in the
extreme blue where the SNR is low.  
The direct images show a compact or stellar object with a close neighbor or
extension to the NE.
CFH12k $Rz'$ imaging (\S\ref{belsample})
suggests that the emission to the NE is a neighbor rather than an extension.

\paragraph{2148.011928}  
This object has one broad line which is identified with \mgii\ at $z=1.063$,
since \civ\ would be missing if the line were \ciii\ and since the line may
show associated absorption.
The object has a high surface brightness core but is clearly extended
to the SE and possibly the NE in the direct images.

\paragraph{1447.090792} 
This object has one broad line which is identified with \mgii\ at $z=1.12$,
since \civ\ would be missing if the line were \ciii.
The line falls atop the 5892\,\AA\ night sky line and thus is partially
interpolated over, but it shows no unusual features
(such as associated absorption) in the 
raw spectrum.
It has a close red neighbor galaxy in the direct images (1447.090780).

\paragraph{1447.150051} 
This object shows one broad line which we identify as \mgii\ at $z=1.165$,
since \civ\ would be missing if the line were \ciii.  
The object appears compact with an extension to the WSW in the $g$ image,
which has the best seeing.

\paragraph{2148.060762}  
This object has one broad line which is assumed to be \ciii\ at $z=1.38$ 
because \OII\ would be missing if the line was \MgII\ at $z=0.624$, 
although that identification cannot be ruled out.  
Possible \AlIII\ emission at the extreme blue end of the spectrum
also supports the \ciii\ redshift.
The object is in the halo of a bright star near the edge of the CCD image,
but it is nonetheless resolved.

\paragraph{1447.162128} 
This object has a blue continuum and one broad line which we identify as 
\ciii\ at $z=1.77$ because of a blue shoulder which matches \AlIII;
however, an identification of \mgii\ at $z=0.89$ cannot be ruled out.
Morphologically, the object is almost certainly unresolved, but is
classified as a probable galaxy.  There is a galaxy of unknown redshift
2\farcs5 to the NE (1447.162145), but the two were separated by PPP so
confusion is unlikely to be responsible for the apparent misclassification.

\paragraph{0223.151505} 
This $z=1.82835$ object shows only one line in the CNOC2 data.
Its identification as \ciii\ comes from a serendipitous spectrum
obtained in 1999 August during a program to measure H$\alpha$ from
a subset of CNOC2 galaxies with known redshifts \citep{mor00}. 
There may be a very faint neighbor or extension to the NW which might be
responsible for this object's classification as a probable galaxy.
However, it is near the corner of a CCD image where image quality is poor
and may in fact be unresolved.

\paragraph{0920.020246}	
This $z=1.94$ object shows \civ, \HeIIsf+\oiii]\,$\lambda$1663, and
\ciii\ emission and possible associated absorption in \civ.
Morphologically the object is classified as unresolved,
but there is faint low surface brightness emission to the S.

This object has been imaged in an ongoing {\em HST} WFPC2 $F814W$ band survey
of CNOC2 galaxy pairs \citep{pat01}.  It is unresolved and
the~low~surface~brightness~emission~is~clearly~a~faint~neighbor~galaxy.

\paragraph{1447.131920} 
This $z=1.965$ object shows strong moderately broad \civ\ emission as well as
\ciii\ and possibly \HeIIsf\ emission.
Morphologically it is a well resolved object.

\paragraph{1447.171440} 
This $z=2.00$ object shows \civ, \HeIIsf+\oiii]\,$\lambda$1663 and 
\ciii\ emission.
It has a neighbor or extension visible to the E in the $R$ and $I$ images
which may be responsible for it being classified as a probable galaxy, 
or it may in fact be very slightly resolved.

\paragraph{1447.132036} 
This $z=2.02$ object shows \civ\ emission with associated absorption
and \ciii\ emission.
The object may be resolved, or it may be unresolved with the faint extension or
envelope visible to the NE responsible for its classification as a probable galaxy.

\paragraph{0223.030993}	 
This $z=2.092$ object shows \civ\ and \ciii\ emission.
Morphologically it is unresolved, but was classified as a probable galaxy in
the summit catalog.  

This object has been imaged in an ongoing {\em HST} WFPC2 $F814W$ band survey
of CNOC2 galaxy pairs \citep{pat01}.  It appears as a very compact object,
consistent with an unresolved nucleus and a faint, 
barely resolved host galaxy.  

\paragraph{0920.092046}  
This object's spectrum shows one broad line with strong associated absorption,
plus several probable absorption lines blueward of the line.
The line could be \mgii\ at $z=0.91$, but the object's relatively red $U-B$ and
very red $B-V$ and $V-R$ colors suggest that \civ\ at $z=2.45$ is more likely
despite the lack of observed \OI, \CII\ or \SiO\ emission. 
Morphologically, on the $R$ image used for classification and mask design it
is situated in a badly defocussed region and is classified as a probable galaxy
in the summit catalog.  It appears resolved on other MOS images and is
clearly resolved in CFH12k $Rz'$ imaging (\S\ref{belsample}).

\paragraph{0223.051932}	 
This $z=2.51$ object shows broad \SiO\ and \civ\ emission 
and narrow absorption in Si\,{\sc iv}\,$\lambda$1400, \CII\ and possibly \OI.
Morphologically it is unresolved, but was classified as a probable galaxy in
the summit catalog.

\paragraph{1447.051899} 
Morphologically this object is classified as a probable galaxy.
It is very compact but resolved, so this is not a misclassification.
There may be faint extended emission to the SE as well.

This object has three definite spectral features:  
a moderately broad line at 5448.7\,\AA,
a broad line at 5042.2\,\AA,
and an absorption trough at $\sim$4833.7\,\AA\ which reaches zero flux.
This trough may be associated with a fourth, less definite spectral feature:
another moderately broad line at 4885.5\,\AA,
or 4866.8\,\AA, if it and the trough are considered a P~Cygni profile.
No two of the observed features form an exact wavelength match
to any pair of common transitions, making the spectrum quite mysterious.

The trough appears to consist of two absorption lines in the 1-D spectrum,
but the evidence for this is weak in the 2-D spectrum, and the noise in the
1-D spectrum is such that the apparent bifurcation could be a fluctuation.
If real, the wavelengths of the two lines (4823.8 \& 4847.5 \AA)
are consistent with either \feii\,$\lambda\lambda$2249.88,2260.78 at $z=1.14$
or \feii\,$\lambda\lambda$2586.65,2600.17 at $z=0.865$.
However, no other lines are identified at either of those redshifts.
We cross-correlated this spectrum with the CNOC2 high-$z$ absorption-line
template (see \S\ref{abssample}) and found $R_{cor}\sim1$ for both possible
\feii\ identifications, indicating no significant cross-correlation peak
at those redshifts.  

The object's red $B-V$ color suggests that the trough is \lya,
although the Lyman decrement does not seem as strong as might be expected
and the object has a noticeably redder spectral index than the
other $z>3$ AGN in our sample (Figure~\ref{f_spectra_bel2}).
If we do assume $z=3.08$, the two definite lines are \NV\ and \CII,
but \OI\ is missing, \SiO\ is missing or weak, 
and \lya\ must be masked by absorption with a complex velocity profile
and partial coverage of the emission region(s).
Missing \lya\ has been seen before in MQS~1114--220 \citep{bh97,bak99},
which has a somewhat similar spectrum if our identification is correct.
Quasars with intrinsically extremely weak \lya\ are known \citep{mcd95,fan99b},
but other lines are similarly weak in those cases.  
Alternatively, we could be seeing the region around \mgii\ in a Low-Ionization
Broad Absorption Line quasar (LoBAL; see \S\ref{Intro}).
Dust, \feii\ absorption, and/or \mgii\ absorption would be required to explain
the red $B-V$ color, which would be unusual but not unique \citep{hal97bal}.
If the trough is \mgii\ absorption at $z=0.74$, the moderately broad line is
\oiii\,$\lambda$3133 but the broad line is unidentified (at 2900\,\AA\ rest).
Thus we consider this identification unlikely, although cross-correlation with 
the composite Large Bright Quasar Survey spectrum \citep[LBQS; ][]{fra91}
yielded its highest value, $R_{cor}$=3.32, at $z$=0.7457$\pm$0.0006.
If the broad line is \mgii\ at $z=0.80$, the moderately broad line is
unidentified but the trough could be \mgii\ blueshifted by 12,000\,\kms.
If the moderately broad line is \mgii\ at $z=0.95$, the broad line could be
\feii\,$\lambda\lambda$2600\,\AA\ in emission, 
and the absorption \feii\,$\lambda\lambda$2600\,\AA\ blueshifted by 15,000 \kms.
We believe this latter identification is unlikely, however, since \feii\ and
\mgii\ have nearly identical ionization levels (as do \feiii\ and \mgiii) and
thus should arise from the same region and should not have different FWHMs,
as observed.
Also, neither of these identifications would explain the 
possible P~Cygni nature of the absorption trough.
A third possibility is that the line and the absorption are both from \civ\ in
an object at $z\sim1.275$, in which case \oiii]\,$\lambda$1663 (but not \HeIIsf)
is seen but considerable dust would have to be invoked to explain the red $B-V$
color.

Given that thousands of galaxy spectra were obtained in CNOC2,
we considered more unusual possibilities.
We ruled out emission from a supernova 
since the spectrum does not match any of the optical spectra
of SNe in \citet{fil97} or any of the UV spectra of SNe in \citet{ctf95}.
We also ruled out the object's being a comet, since it did not move between
two images taken a year apart.  
Our direct images rule out two superimposed broad-line objects,
but not the possibility of one broad-line object lensing another. 
However, as mentioned above, no two of the observed features exactly match any
common pair of emission lines, which would be expected in the lensing scenario.

We have tentatively adopted $z=3.08$ since it seems the least unlikely redshift.
We note that \citet{hat00} calculate a photometric redshift $z_{phot}\simeq3.71$
for this object, consistent with $z=3.08$ given their observed dispersion in
$z_{phot}$ vs. $z_{spec}$ for the CNOC2 AGN with firm redshifts.

\paragraph{2148.112371}  
The redshift of this $z=3.157$ object comes from \lya\ and \nv\ emission.
No \OI\ emission is detected and \CII\ and \SiO\ fall atop the 5577\,\AA\ and
5892\,\AA\ night sky lines, respectively.
This unresolved object was added to its spectroscopic mask by hand to fill an
empty area.

\paragraph{1447.020157} 
This $z=3.17$ object shows Ly$\alpha$ and probable weak \SiO\ emission plus 
Ly$\alpha$ forest absorption which causes a change in the continuum slope 
across the line.  There are also
two narrow absorption lines at $\sim$5442\,\AA\ and $\sim$5759\,\AA\ observed
which are obvious in the 2-D spectroscopic images as well as the 1-D spectrum.
Morphologically the object is unresolved and unremarkable.

\paragraph{2148.171412}  
The redshift of this $z=3.19543$ object comes from 
strong, narrow \lya\ emission and a shoulder of \nv\ emission.  
The redshift is corroborated by the change in the continuum slope across
the line due to \lya\ forest absorption, and by the object's broadband colors.
No \OI\ emission is detected and \CII\ and \SiO\ fall atop the 5577\,\AA\ and
5892\,\AA\ night sky lines, respectively.  
The equivalent width 
for the latter line was measured from the unweighted extraction,
which is not interpolated across these lines.
This compact object falls partially atop a bleeding column from a nearby bright
star in the $R$ image, but it is classified as a galaxy in the $g$ image, so we
consider the classification secure.

\paragraph{0223.151510} 
This $z=3.5$ object shows \lya\ and \nv\ emission.
Flux is seen redward but not blueward of the line in the 2-D spectroscopic
images, but the SNR is not high enough to measure a flux decrement across
the line with any accuracy.
It is the easternmost member of a pair of compact objects, with a brighter 
and bluer neighbor located $\sim$1\farcs3 to the WNW 
as measured in the $g$ image, which has the best seeing.  
The objects were not separated by PPP and were classified as a single galaxy.
The AGN is consistent with being unresolved, but the neighbor does appear
resolved in the $g$ image at least.
The neighbor is brighter than the AGN in the direct images but fainter in the
2-D spectroscopic images, possibly due to the slit positioning.
No redshift was measured for the neighbor galaxy due to its very faint spectrum.

\paragraph{0223.010767}	
This $z=4.00$ object shows \lyb+\ovi, \lya, and \nv\ emission, 
plus absorption from the Lyman forest blueward of \lya.
This object 
is morphologically classified as a probable galaxy, 
possibly due to confusion from a galaxy of unknown redshift (0223.010775)
located 2\farcs5 N of this object.  
The object is consistent with being unresolved 
(the $R$ image has a slightly elongated PSF due to poor focusing).

\paragraph{0223.011745}  
The one broad line visible in this object's spectrum is identified as
\lyb+\ovi\ at $z=4.67$ for several reasons.
There is an absorption system just blueward of the line for which
matching Ly$\gamma$ is seen.  
The spectrum is consistent within the flux calibration uncertainties with no
flux below 5000\,\AA, 
which can be explained by a Lyman limit system at $z=4.48$.  
There are also numerous absorption features present all the way from 5000 to
6300\,\AA, consistent with the \lya\ forest.
Last but not least, the object's $B-R$ color is very red (\S\ref{Colors}), much
redder than normal stars or galaxies but exactly as expected for a $z>4$ quasar.
The equivalent width of the line is high for \lyb+\ovi\ but would be
low for Ly$\alpha$.
Also, if the emission line is Ly$\alpha$ at $z=3.79$, it is difficult to
explain the lack of appreciable $B$-band flux in the spectrum or photometry.
Nonetheless, a spectrum extending farther to the red is needed to confirm
our assumed redshift, which is the highest in the sample.
Morphologically the object is classified as stellar in the final catalog, 
but it was targeted for spectroscopy because it was classified as a probable
galaxy in the summit catalog due to the image being slightly out of focus
at this location on the CCD.  
There is a $z=0.45777$ galaxy 3\arcsec\ to the SE (0223.011732).

\section{[Ne\,{\sc v}] AGN}	\label{nevsample}

Given the 97.02~eV ionization potential of \nev,
the presence of \NeVb\ or \NeVa\ emission should be a
robust signature of AGN for objects where \mgii\ is absent or not observable.
\NeVa\ arises from the same excited level as \NeVb\ but 
with a relative intensity of 0.359 \citep{row93}.
A \nev-selected AGN sample should also suffer little contamination:
\citet{bpt81} found that no \hii\ region in their large literature sample
showed detectable \NeVb,
and \citet{skc95} detected \NeVb\ only among the Seyfert~2 galaxies in their
sample of low-redshift Seyfert~2 and star-forming galaxies.
In our sample, \nev\ emission is seen along with broad \MgII\ in
2148.102026, 0920.182673, and possibly 2148.151660
and is detected in eight other objects as detailed below.

To establish the sensitivity of our spectra to \nev\ emission, 
we force-fitted an unresolved Gaussian at the position of \NeVb\ to
the spectrum of each of the $\sim$4000 
$z>0.281$ objects in which \NeVb\ falls within our wavelength coverage.
The resulting distribution of equivalent widths reflects the noise
in our spectra at rest-frame 3426\,\AA.
The \NeVb\ line with the smallest observed rest-frame equivalent width (\ew)
would have been detectable in only 31\% of $z>0.281$ objects, while the
\NeVb\ line with the largest observed \ew\ would have been detectable in 93\%.
The average \nev\ detectability, 
as measured by our detections, 
is 71$\pm$20\%.  

\subsection{Notes on Individual [Ne\,{\sc v}] AGN}	\label{nevnotes}

\paragraph{2148.170418}
This $z=0.32$ object shows \NeVb, \oii\ and \NeIIIa\ emission
plus Ca~H+K, CN, and Balmer absorption lines from H$\delta$ to H10 (H$\gamma$
appears to be in emission, albeit at the noisy red end of the spectrum).
Morphologically the object appears resolved, but the PSF is rather poor.

\paragraph{0223.040997} 
This $z=0.39868$ object shows \NeVa, \oii\ and \NeIIIa\ emission,
plus Ca~H+K and CN absorption.
Morphologically, this object is clearly resolved.  It has a diffuse blue
$z=0.39897$ emission-line companion (0223.040989) 2\farcs1 to the S
and a very faint extension or neighbor slightly W of N.

\paragraph{1447.150676} 
This $z=0.47$ object shows \NeVb, \NeVa, \oii\ and \NeIIIa\ emission,
plus Ca~K absorption.  
Morphologically, the object is resolved and has an absorption-line companion
(1447.150680) at $z=0.46692$ 4\farcs5 to the ENE.

\paragraph{1447.080513}
This $z=0.48$ object shows \NeVb, \NeVa, \oii\ and \NeIIIa\ emission 
plus Ca~H+K, CN, and H$\delta$ absorption.
A zeroth-order region in the blue has been interpolated over,
and there is a spurious line in the extreme blue.
Morphologically, this object is compact but resolved.

\paragraph{1447.140343}
This $z=0.51$ object shows \NeVb, \NeVa, \oii\ and \NeIIIa\ emission 
(the latter two of which overlap excised night sky line regions) plus Ca~H
and possibly CN and H$\delta$ absorption (Ca~K lies atop a night sky line).
Morphologically, this galaxy is compact but resolved and extended E-W.

\paragraph{0223.130940}  
The object appears unresolved in the direct images and is
classified as an AGN based on the presence of \NeVb.
The redshift of 0.53942 is from cross-correlation with the emission-line
template, and is based primarily on narrow \oii\ emission and Ca~K absorption,
although \NeIIIa\ is also present.
CN absorption falls atop the 5892\,\AA\ night sky line
and Ca~H absorption is absent.  However, there is a suggestion of emission from 
H$\zeta$ or \hei\,3889\,\AA\ which might suggest that Ca~H is filled in by H$\epsilon$
emission.  
With this redshift, 
the moderately broad \oiii\,$\lambda$3133 and \NeVb\ lines and the narrow 
\NeIIIa\ line appear blueshifted by $-$370$\pm$120\,\kms.  
This object also exhibits broad \oiii\,$\lambda$3133 absorption.
This is to our knowledge the first recognized case of such absorption.
However, it may also be present in the spectra of the 
LoBAL 
quasar PG~1700+518 presented by \citet{pet85} and
\citet{ogl99}, although it is difficult to separate from \feii\ emission at 
$<$3000\,\AA\ and at 3130--3320\,\AA.
\oiii\,$\lambda$3133 does satisfy the criterion for a transition to show broad
absorption outlined in Eq.~1 of \citet{kor92}.
One lingering concern with our adopted redshift
is that \NeVa\ is not seen at the expected level,
a nondetection significant at $\sim7\sigma$ given the observed noise level.
However, \NeVb\ appears stronger in the 1-D spectrum than in the 2-D spectral
images, so if it is spuriously strong in the 1-D spectrum then the limit on the
\NeVa\ emission might be consistent with the expected ratio of 0.359.
	Alternatively, the detailed continuum shape could conceivably 
	conspire to help hide the \NeVa\ emission.
	Another concern with the redshift is that there is a possible 
	unidentified weak emission line at 3657\,\AA\ rest (5630\,\AA\ 
							    observed),
	but it is not obvious in the 2-D image so it may be spurious.

This object has been imaged in an ongoing {\em HST} WFPC2 $F814W$ band survey
of CNOC2 galaxy pairs \citep{pat01}.  It appears as a bulge-dominated galaxy
with a nuclear point source, consistent with the presence of an AGN.
However, a cuspy bulge rather than a nuclear point source cannot be ruled 
out without detailed modeling.

\paragraph{0920.161307} 
This $z=0.56147$ object shows \NeVa, \NeVb, \OII, \NeIIIa\ and possibly
\oiii\,$\lambda$3133 emission, plus Ca~K absorption.
Morphologically, it is compact but resolved and has an extension or close
neighbor to the SW.  
CFH12k $Rz'$ imaging (\S\ref{belsample})
also reveals faint, diffuse emission to the SE of both objects,

\paragraph{0920.020116}	
This $z=0.56225$ object shows narrow \OII, \NeIIIa, \NeVb\ and possibly
\oiii\,$\lambda$3133 emission.  \NeVa\ is not seen at the expected level, 
but the nondetection is consistent with the noise level.
There is also a poorly interpolated zeroth-order region in the blue.
Morphologically it is a resolved but compact, high surface brightness galaxy
on both MOS and CFH12k imaging (\S\ref{belsample}).

\section{Absorption-Line AGN Candidates}	\label{abssample}
Some AGN are known to show strong absorption and little or no line emission
in the rest-frame UV \citep{smi95,ega96,bec97}.  
The CNOC2 spectra can detect absorption-line AGN at $0.58<z<1.24$ via \MgII,
at $0.70<z<1.41$ via \feii\,$\lambda\lambda$2600 and at $0.89<z<1.67$ via
\feii\,$\lambda$2344 
(see \S\ref{absnotes} for the definition of this line notation).

Visual inspection yielded 
twelve candidate absorption-line AGN.  
Cross-correlation allows a better estimate of the robustness and completeness
of the detections, but we could find no template in the literature 
which matched these objects' spectra and had acceptable SNR.
Thus we constructed a template high-$z$ absorption-line spectrum using the six
candidates judged to be the most reliable visual detections 
($R_{cor}\equiv5$ in Table~\ref{tab_spec_abs}).
Each spectrum was shifted to its rest frame, 
linearly interpolated to a common wavelength scale, 
normalized by a single third order spline fit to the entire spectrum
after a maximum of 10 iterations of $\pm$2$\sigma$ pixel rejection,
and scaled to unity in the common wavelength range
2525--2702\,\AA\ (excluding 2575--2610\,\AA).
The spectra were averaged at all wavelengths where at least two objects 
overlapped and again normalized by a single third order spline.
New redshifts were calculated for the input spectra and the procedure was
repeated twice to produce the final high-$z$ absorption-line template.
All $\sim$4500 unclassified, mostly very low SNR CNOC2 spectra and all visually
selected absorption-line AGN were cross-correlated with this template.
By inspecting the 5\% of objects with the highest $R_{cor}$ ($R_{cor}\geq3.15$),
we recovered all but the weakest visually selected absorption-line candidate and
added four new objects (0920.060299, 0920.110823, 1447.020849 and 1447.111559),
for a total of fifteen candidate absorption-line AGN at $0.73 < z < 1.33$.
To be sure we were not missing any $z > 1.33$ absorption line AGN candidates,
or misidentifying such objects as lower-$z$ objects, we used the lensed $z=2.72$
galaxy MS 1512--cB58 \citep{yea96,pea00} as a template for cross-correlation of
all candidate $z < 1.33$ absorption-line AGN and all unclassified but high SNR
CNOC2 spectra.  No additional candidates with $R_{cor}\geq3.15$ were identified,
and none of the candidate absorption-line AGN had higher $R_{cor}$ values when
cross-correlated with this template.

Only one of these candidate absorption line AGN (2148.050488)
has been confirmed as a broad-line AGN via IR spectroscopy (\S\ref{irspec}).
The rest remain only candidates (see \S\ref{morphs}),
since 3--8\% of CNOC2 galaxies at $z>0.58$ show \MgII\ absorption
and since lines of the observed strengths can be produced in starburst 
galaxies \citep{skc95}, as discussed in some detail by \citet{plo98}.

\subsection{Notes on Individual Absorption-Line AGN Candidates}	\label{absnotes}

In the following notes
\mgii\ denotes lines at 2796.35 and 2803.53\,\AA,
\feii\,2255 lines at 2249.88 and 2260.78\,\AA,
\feii\,2380 lines at 2374.46 and 2382.77\,\AA,
and \feii\,2600 lines at 2586.65 and 2600.17\,\AA\
(possibly confused with Mn\,{\sc ii} lines at 2576.88, 2594.50 and 2606.46\,\AA).

\paragraph{0920.110823} 
This $z=0.73337$ object shows \feii\,2600 and \mgii\ absorption.
Morphologically it is compact but extended and has a neighbor (110820)
1\farcs5 S of E.

\paragraph{2148.162196} 
This $z=0.74842$ object shows 
\feii\,2600 and \mgii\ absorption.
The apparent emission just redward of the 5577\,\AA\ night sky line is spurious,
and no other absorption lines are seen in both independent spectra of the
object.  Morphologically the object is asymmetrical and slightly elongated.

\paragraph{2148.150370} 
This $z=0.75621$ object shows \feii\,2600, metastable \feii*\,$\lambda$2747
and \mgii\ absorption.
The apparent broad emission surrounding the \feii\,2600 emission is spurious.
Morphologically the object consists of a nucleus in a relatively
large elliptical halo, and it has a very faint close neighbor (2148.150382).

\paragraph{0920.060299} 
This $z=0.79999$ object shows very narrow \mgii\ and possibly 
\feii\,2600\,\AA\ absorption.  Morphologically it is clearly resolved, with
an extension to the W reminiscent of a spiral arm or possibly a tidal tail.

\paragraph{0223.190667} 
This $z=0.847$ object shows \mgii\ and 
\feii\,2600\,\AA\ absorption.  
Morphologically it is clearly a galaxy with an extension or neighbor to the NW.

\paragraph{2148.121113} 
This $z=0.88595$ object shows \mgii\ and weak \feii\,2600 absorption.
Morphologically the object is clearly resolved.

\paragraph{0920.101758} 
This $z=0.89979$ object shows \mgii\ and 
\feii\ 2344, 2380, 
and 2600\,\AA\ absorption.
Morphologically it is a compact galaxy extended in the E-W direction.

\paragraph{2148.130511} 
This $z=0.91352$ object shows \mgii\ and \feii\ 2380 
and 2600 absorption,
and possibly \feii\,2600 emission with a P~Cygni profile.
Morphologically the object is clearly resolved and slightly asymmetrical,
and there is a red galaxy (2148.130526) located several arcseconds to the north.

\paragraph{1447.140898} 
This $z=1.01769$ object shows \mgii\ and \feii\ 2344, 2380 
and 2600\,\AA\ absorption.
It has a very blue continuum which suggests it is an AGN rather
than a starburst galaxy.  
However, the direct images show a compact but extended, elongated galaxy 
with a very close faint neighbor.

\paragraph{0223.071845} 
This $z=1.06096$ object shows \mgii\ and \feii\ 2344, 2380 
and 2600\,\AA\ absorption, as well as numerous possible unidentified lines
blueward of \feii\,2344\,\AA.  The object has a blue continuum and tentative
narrow \mgii\ and \feii\,2600\,\AA\ emission, which suggests it is an AGN.
Morphologically the object is compact but resolved, with a galaxy of unknown
redshift (0223.071841) located 4\farcs2 slightly S of W.

\paragraph{1447.150061} 
This $z=1.07162$ object shows \mgii\ and \feii\ 2344 
and 2600\,\AA\ absorption.
The emission observed at 5725\,\AA\ is spurious.
This unresolved object was added to its spectroscopic mask by hand to fill an
empty area.

\paragraph{1447.020849} 
This $z=1.07458$ object shows \mgii\ and \feii\ 2255, 2344, 2380 and 
2600\,\AA\ absorption.  Morphologically the object is compact but resolved,
with a faint uncatalogued neighbor located a few arcseconds slightly N of W.

\paragraph{1447.091045} 
This $z=1.11503$ object shows \feii\ 2344, 2380 
and 2600\,\AA\ absorption.  On the direct images the object is compact 
but resolved, possibly a double with a small neighbor to the W.

\paragraph{1447.111559} 
This $z=1.17736$ object shows \mgii\ and \feii\ 2344, 2380
and 2600\,\AA\ absorption.  Morphologically the object is compact but resolved,
with~a~faint~neighbor~(1447.111562)~located~2\farcs5~slightly~N~of~E.

\paragraph{2148.050488} 
This object shows strong \feii\ 2344, 2380 
absorption at $z=1.32701$ and possible \CIII\ emission at the extreme blue end
of the spectrum.  We consider it our only confirmed absorption-line AGN,
as an infrared spectrum reveals broad H$\alpha$ emission (see \S\ref{irspec}).
In the direct images the object is unresolved but has a close neighbor
(2148.050494).

\subsection{Infrared Spectroscopy of 2148.050488}	\label{irspec}

\begin{figure}
\epsscale{0.65}  
\plotone{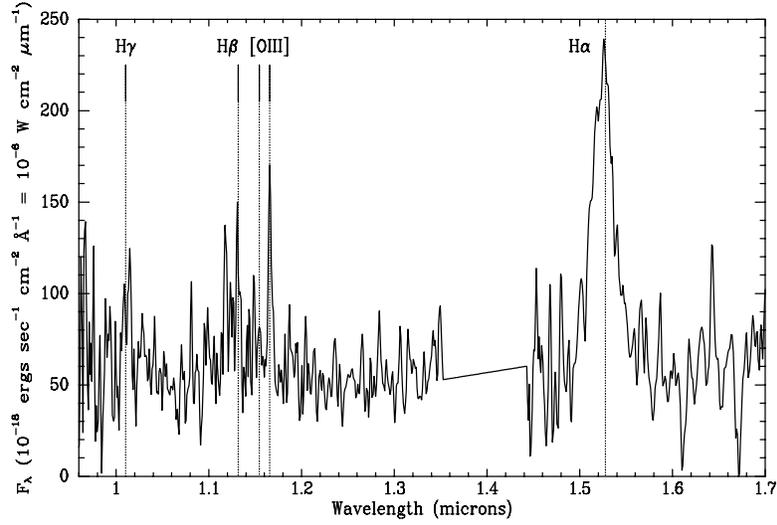} 
\caption[]{
\singlespace
Spectrum of 2148.050488 from 0.96--1.7\,\micron.  The low signal-to-noise
region between the $J$ and $H$ bands has been interpolated over.
Dotted lines are drawn at the expected wavelengths of major emission lines
at the systemic redshift of $z=1.328$.
}\label{f_jh2148.050488}
\end{figure}

The CNOC2 absorption-line AGN candidate 2148.050488 
was observed by R. Elston on UT 1997 September 13 with
the CTIO 4m and Infra-Red Spectrograph \citep{irs90}.  In cross-dispersed mode, 
IRS uses a 256x256 InSb array to obtain a spectrum with resolution $R$=560 from 
0.96--2.4\,\micron\ in a single exposure.  
Twenty-two 200-second exposures were obtained, 
and standard observing and reduction procedures were used 
to produce the final spectrum, part of which is shown in 
Figure~\ref{f_jh2148.050488}.
H$\alpha$ is clearly detected, with 
FWHM$\sim$1.16~10$^4$\,\kms\ and 
FWZI$\sim$3.0~10$^4$\,\kms,  
broader than average for AGN but within the observed range \citep{pet97}.
\OIII\,$\lambda\lambda$4959,5007, H$\beta$, and possibly H$\gamma$
are also detected.
The \OIII\ redshift is 1.328$\pm$0.001.
Relative to this systemic $z$, the \feii\ absorption in the discovery spectrum
($z$=1.32701$\pm$0.00013) is blueshifted by --220$\pm$300\,\kms,
and H$\alpha$ is blueshifted by --980$\pm$330\,\kms.
This is an extreme but not unprecendented blueshift for H$\alpha$ \citep{mci99}.

The procedure outlined in \S\ref{EWs} was used to find the \ew\ for each line
in the IR spectrum.\footnote{This procedure uses wavelength ranges defined for
each line in the average LBQS spectrum \citep{fra91}.  
The wavelength range for H$\beta$ includes \OIII\,$\lambda\lambda$4959,5007,
so the final \ew(H$\beta$) is the raw measurement minus \ew(\OIII).
The LBQS did not observe H$\alpha$, so for it we used a wavelength range
corresponding to the velocity range used for H$\beta$.
}
The results are given in Table~\ref{tab_spec_abs}.
The Balmer decrement in 2148.050488 is 3.34$\pm$0.81, 
consistent with the range of 4$-$6 seen in typical quasars \citep{pet97}
and much smaller than the values of 10$-$13 found for the
narrow-absorption-line AGN Hawaii~167 and Q~0059-2735 by \citet{ega96}.
H$\beta$ appears to have a smaller FWHM than H$\alpha$,
which would mean that the Balmer decrement increases away from the line center,
the opposite of what is usually seen.
A spectrum with higher SNR than ours is required to confirm this.

The broad H$\alpha$ line confirms 2148.050488 as a broad emission line AGN,
albeit with an unusually large blueshift and FWHM for H$\alpha$,
but the normal Balmer decrement shows that it is not an extremely reddened AGN.
Optical spectra with wider wavelength coverage could determine if it is
a BAL, a LoBAL, of a normal broad emission line AGN with
strong associated absorption in \feii\ and perhaps other species.

\section{Discussion} 	\label{Discussion}

\subsection{Morphologies} 	\label{morphs}

The CNOC2 AGN sample is the largest sample to date of distant AGN selected via
medium-resolution optical spectroscopy, without reference to broadband color.
The sample is thus limited primarily by magnitude and line detectability.
The CNOC2 AGN sample is also unique in that only objects initially classified
as galaxies or probable galaxies were targeted for spectroscopy,
with the exception of three objects added by hand to fill empty areas
on the masks (1447.081984, 1447.150061, and 2148.112371)
and a handful of serendipitously observed unresolved objects.
Thus this sample is not biased against
objects with luminous host galaxies or with close neighbors,
except for possible instances of extreme extinction in the latter case.
The latter case means the sample is also not strongly biased against AGN with
redder colors due to dust reddening by foreground neighbors.
In this section we consider in more detail the morphologies of these AGN
and the implications thereof.

\begin{figure}
\epsscale{0.65}
\plotone{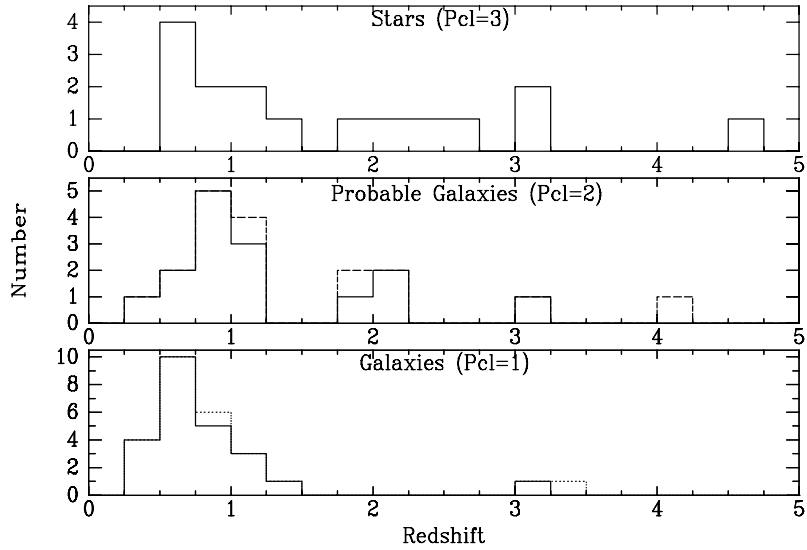} 
\caption[]{
\singlespace
Distribution of automated PPP morphological classifications with redshift.
The dashed entries (middle panel) show the two probable misclassifications
of stellar objects, and the dotted entries (lower panel) the two definite 
misclassifications of stellar objects.  Misclassifications are typically 
caused by very close neighbor galaxies not separated by PPP.
}\label{f_morphs}
\end{figure}

Table~\ref{tab_phot} gives the PPP morphological class (Pcl) of each object.
PPP automatically classifies objects as either galaxies (Pcl=1),
probable galaxies (Pcl=2), stars (Pcl=3), or saturated stars (Pcl=4).
Among the 46 emission line selected AGN, we find 
two definite (0223.151510, 0920.080251), 
two probable (0223.010767, 1447.162128), and 
four possible (0223.151505, 1447.161148, 1447.171440, 2148.102026)
cases of classification errors where an unresolved object (star) was mistakenly
classified as resolved (galaxy or probable galaxy) in the final catalog, 
or vice versa in one case.
After correcting for definite and probable classification errors,
the 46 emission line AGN are distributed among stars, probable galaxies, and
galaxies in the ratio 1.4:1:1.15.
However, the 15 absorption line objects 
are distributed among these categories in the ratio 1:1.5:5.
This is consistent with the expected contamination of the
absorption-line objects with star-forming galaxies (\S\ref{abssample}).
Figure~\ref{f_morphs} shows the distribution of automated PPP morphological
classifications with redshift.  
Objects with possible or probable errors are plotted with different line types.
AGN can be unresolved at redshifts as low as $z=0.5$,
but can also be resolved by PPP at redshifts as high as $z\simeq3$, 
even with coarse sampling of the subarcsecond to arcsecond PSF achieved at CFHT.

The resolved and unresolved CNOC2 AGN are expected to be
representative samples of each of those AGN types.
However, since spectra were obtained for many more resolved than unresolved
objects, the fraction of resolved AGN in the CNOC2 sample is higher than among
AGN as a whole (\S\ref{Density}).  This may mean that the sample is biased
toward AGN with 
host galaxies more luminous than average for those AGN luminosities.
To quantify such a bias would require targeted spectroscopy to assemble a large
sample of AGN from the CNOC2 photometric catalog for comparison with this one.
Finally, with only two definite cases of unresolved objects misclassified as
resolved due to overlooked close neighbors, the CNOC2 AGN sample is probably 
not strongly affected by any biases introduced by such neighbor galaxies,
except perhaps for microlensing by stars in their halos \citep{can81}.

\subsection{Broad-Band Colors and Color Selection} 	\label{Colors}

As discussed in \S\ref{Intro}, the efficiency of color selection of quasars
is of interest in determining the true and relative space densities of all
quasars and of various quasar subclasses.  Thus in this section we compare
the colors of the CNOC2 AGN with the Deep Multicolor Survey (DMS) quasar
sample \citep{hal96dms1,hal96dms2,hal96dms2err,ken97dms3,osm98dms4}.

\begin{figure}
\epsscale{1.0}
\plotone{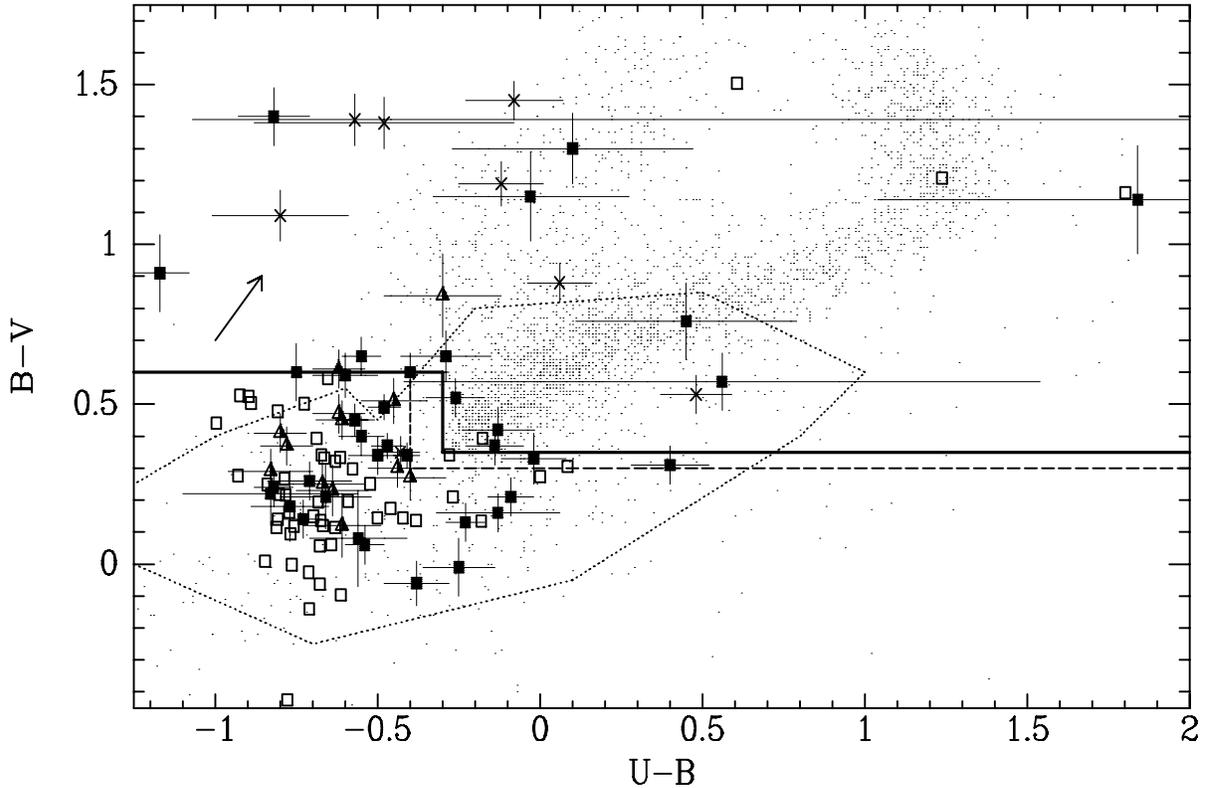} 
\caption[]{
\singlespace
$U-B$ vs. $B-V$ color-color diagram showing the 59 CNOC2 AGN and AGN candidates
with $UBV$ data, plus the 53 color-selected Deep Multicolor Survey (DMS) AGN
\citep{osm98dms4}.  
Filled squares are broad-emission-line AGN,
crosses are \nev\ AGN;
filled and half-filled triangles are confirmed and candidate absorption-line
AGN, respectively;
open squares are AGN from the DMS (error bars not plotted);
and points are all $\sim$3500 $B\leq21$ objects in the CNOC2 catalog.
The catalog has not been checked for saturated stars or other erroneous colors.
One \nev-selected AGN has only a $2\sigma$ lower limit to its $U-B$ color.
The arrow shows the reddening vector for $E(B-V)=0.2$ \citep{mb81}.
The dotted line encloses the region occupied by the simulated $z<3$ quasars of
\citet{hal96dms2}.
The thick solid and dashed lines show the bright ($B<21$) and faint ($21<B<22$)
selection criteria of \citet{hal96dms2}; objects blueward of the lines in
$B-V$ in those magnitude ranges were considered quasar candidates.
}\label{f_ubbv}
\end{figure}

Figure~\ref{f_ubbv} shows the $U-B$ vs. $B-V$ colors of
the 59 CNOC2 AGN and AGN candidates with $UBV$ data (various symbols),
the 53 color-selected DMS AGN (open squares),
and all $B\leq21$ objects in the CNOC2 photometric catalog (dots).
The high galactic latitude stellar locus extending from F stars at (--0.2,0.4)
to M stars at (1,1.6) is visible, along with the locus of galaxies located
blueward of the stellar locus in $U-B$ (at given $B-V$) and redward of it in
$B-V$ (at given $U-B$) due to the composite stellar populations of galaxies.
The dotted line encloses the region occupied by the simulated $z<3$ quasars of
\citet{hal96dms2}.
The thick solid and dashed lines show the bright and faint selection criteria
of \citet{hal96dms2}; objects blueward of the lines in $B-V$ at $B<21$ and
$21<B<22$, respectively, were selected as candidate $z<3$ quasars.
The criteria do not reach as red in $B-V$ at $U-B\gtrsim0.35$ as at
$U-B\lesssim0.35$ because of the need to avoid the stellar locus, 
and the criteria do not probe as close to the stellar locus at fainter
magnitudes due to the spreading of the locus from photometric errors.

\begin{figure}
\epsscale{1.0}	
\plotone{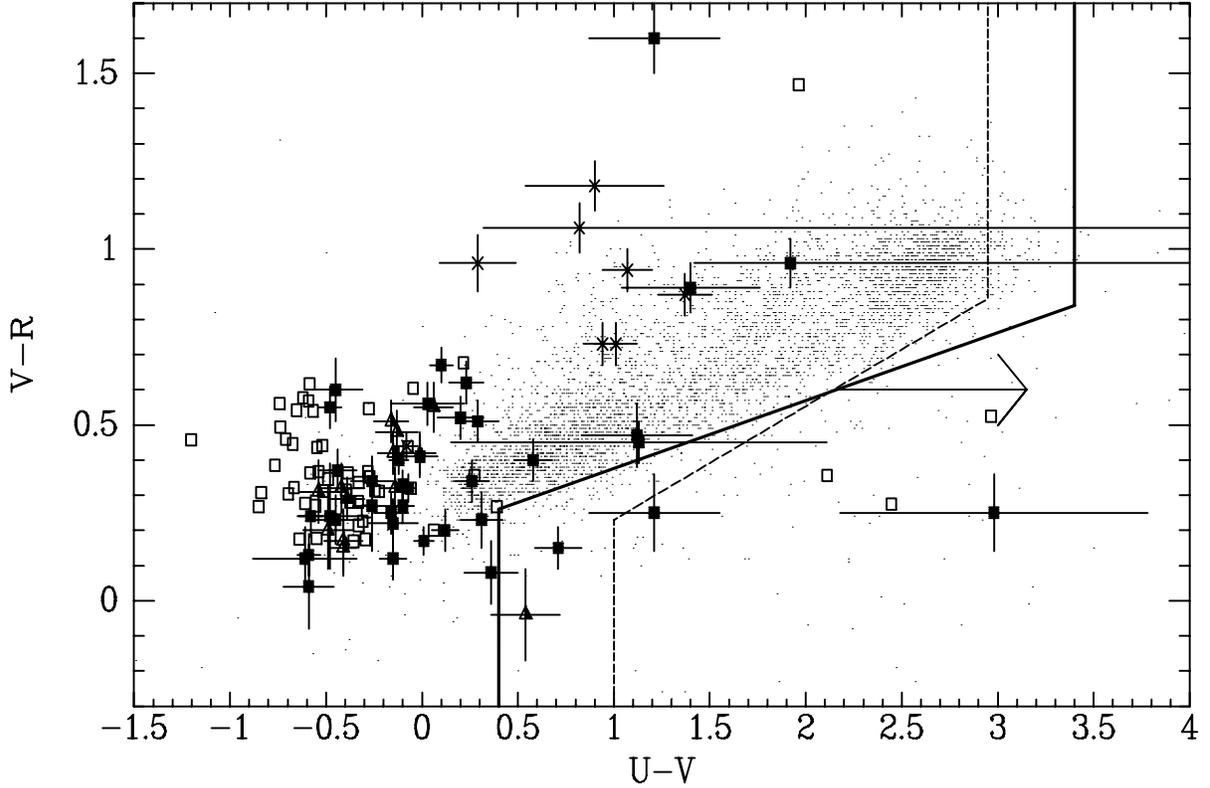} 
\caption[]{
\singlespace
$U-V$ vs. $V-R$ color-color diagram showing the 59 CNOC2 AGN and AGN
candidates with $UVR$ data, plus the 53 color-selected DMS AGN.
Points are all $\sim$4850 $17<V<20.5$ objects in the CNOC2 photometric catalog.
Other symbols are the same as in Figure~\ref{f_ubbv}.
Two CNOC2 AGN have only $2\sigma$ lower limits to their $U-V$ colors.
The thick solid and dashed lines show the bright ($17<V<20.5$) and faint
($20.5<V<22$) $z\gtrsim3$ quasar selection criteria of \citet{hal96dms2}; 
objects in that magnitude ranges {\em redward} of the lines in $U-V$ 
(as indicated by the large arrow)
{\em and} in $B-V$ (Figure~\ref{f_bvvr}) were considered quasar candidates.
}\label{f_uvvr}
\end{figure}

\begin{figure}
\plotone{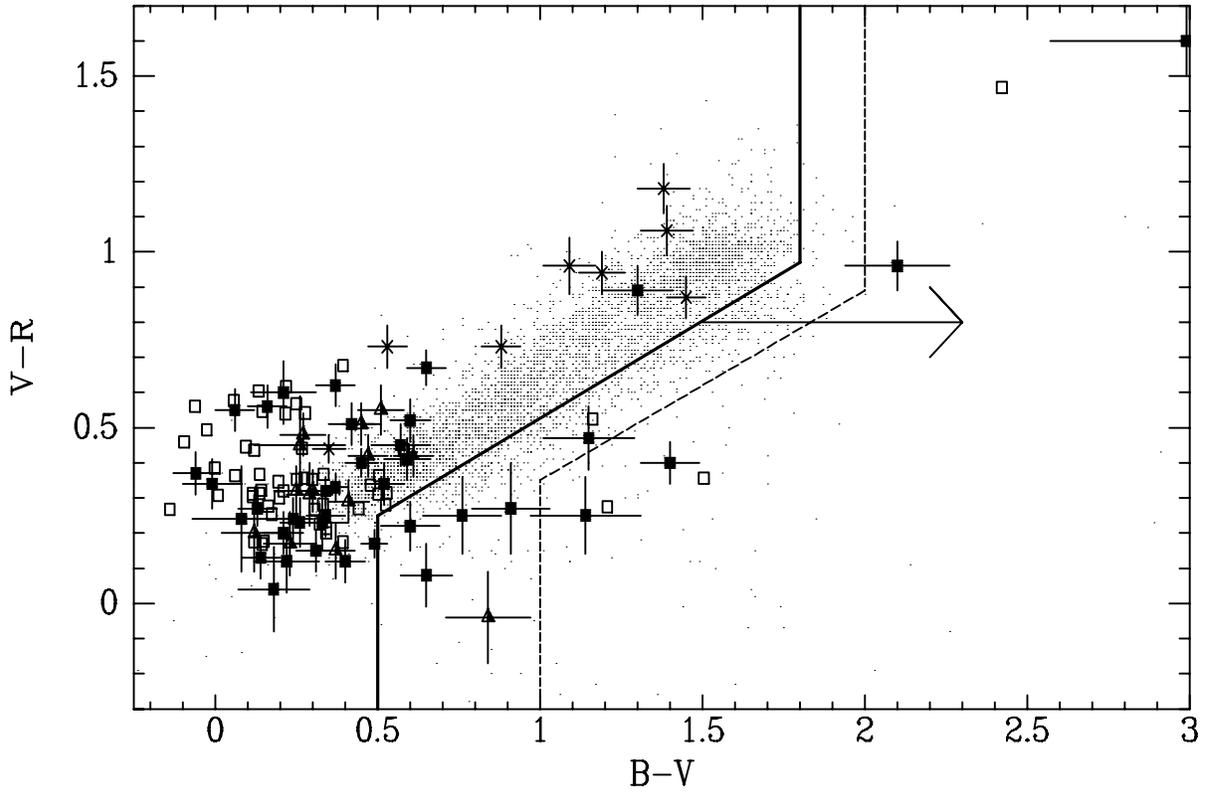} 
\caption[]{
\singlespace
$B-V$ vs. $V-R$ color-color diagram showing the 60 CNOC2 AGN and AGN
candidates with $BVR$ data, plus the 53 color-selected DMS AGN.  
Points are all $\sim$4850 $17<V<20.5$ objects in the CNOC2 photometric catalog.
Other symbols are the same as in Figure~\ref{f_ubbv}.
The thick solid and dashed lines show the bright ($17<V<20.5$) and faint
($20.5<V<22$) $z\gtrsim3$ quasar selection criteria of \citet{hal96dms2};
objects in that magnitude ranges {\em redward} of the lines in $B-V$ 
(as indicated by the large arrow)
{\em and} in $U-V$ (Figure~\ref{f_uvvr}) were considered quasar candidates.
}\label{f_bvvr}
\end{figure}

\begin{figure}
\plotone{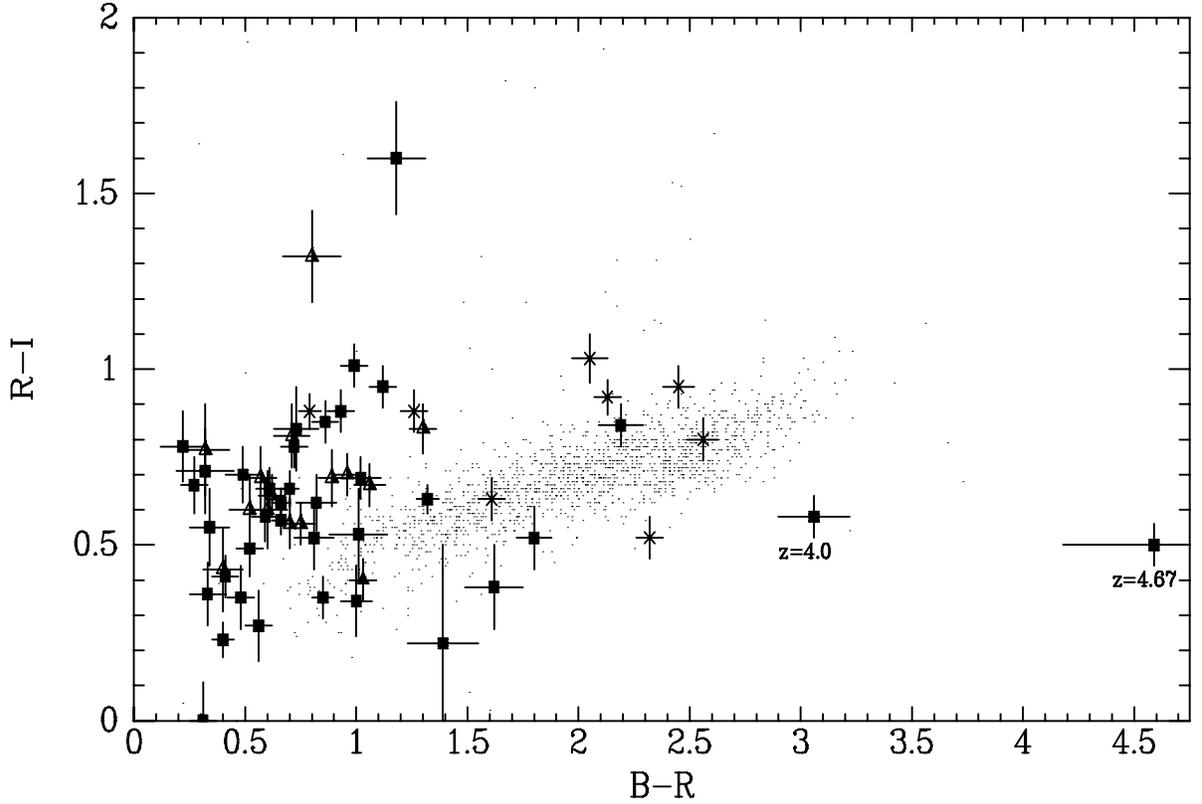} 
\caption[]{
\singlespace
$B-R$ vs. $R-I$ color-color diagram showing the 60 CNOC2 AGN and AGN candidates
with $BRI$ data.  
The CNOC2 AGN with $R-I=1.6$ has a spuriously red $I$ magnitude due to a nearby
bright star.
Points are all $\sim$1700 $R\leq19.7$ objects in the CNOC2 photometric catalog.
Other symbols are the same as in Figure~\ref{f_ubbv}.
The $z=4.00$ object 0223.010767 and the $z=4.67$ object 0223.011745 
are marked with their redshifts.
No DMS AGN or selection criteria lines are included because the DMS did not use
a standard $I$ filter.
}\label{f_brri}
\end{figure}

Figures~\ref{f_uvvr}--\ref{f_bvvr} and Figure~\ref{f_brri}
show similar color-color diagrams for use in selecting
$3\lesssim z \lesssim4$ and $z>4$ quasars, respectively.
The bright and faint selection criterion lines cross in the $U-V$/$V-R$ diagram 
(Figure~\ref{f_uvvr}) because the
faint criteria were nearly reaching the limiting depth of the DMS $U$ data.
Some adjustment to the DMS bright criteria in the $B-V$/$V-R$ diagram 
(Figure~\ref{f_bvvr}) is probably warranted to avoid probing too close
to the locus of stars and galaxies in the CNOC2 colors.
No DMS selection criteria are drawn in the $B-R$/$R-I$ diagram
(Figure~\ref{f_brri}) because the DMS did not use a standard $I$ filter.
A simple selection criterion for $z>4$ quasar candidates is $B-R>3.3$,
which recovers the $z=4.67$ object 0223.011745 with $B-R=4.6$.
Through extension to less red $B-R$ but also blue $R-I$ the selection becomes
sensitive down to $z\sim3.3$ and more efficient at $z>4$.
For example, the $z=4.00$ object 0223.010767 at $B-R=3.1$ would be recovered.

As a simple example of AGN color selection, 
we use the bright-magnitude selection criteria of \citet{hal96dms2}.
With these criteria, 42 of 59 CNOC2 AGN and AGN candidates with $U$ data
would qualify as color-selected AGN candidates (71\%).
Two of the seventeen missed objects
would be selected if the $U-B$/$B-V$ selection criterion
extended just 0\fm1 redder in $B-V$ at $U-B<-0.3$.
Four others would be selected if they were required to be outliers in just
the $B-V$/$V-R$ diagram rather than both it and the $U-V$/$V-R$ diagram.
(The $B-V$/$V-R$ selection criteria of \citet{hal96dms2} probe too close to the
  stellar locus to be efficient, but the AGN in question are sufficiently
  far from the locus to be recovered using reasonable criteria.)
With those revised criteria, 81$\pm$6\% of the CNOC2 AGN
would qualify as color-selected AGN candidates (48 of 59 with $UBV$ data).
We have assumed that the spread in the stellar locus is negligible at the
magnitudes of interest, so 
this will be an upper limit to the efficiency of color selection.
Variability has not been accounted for either, but it should be a
small effect since most images of a given CNOC2 field were taken 
within a few days of each other.
Also, \citet{who91} find that variability generally increases detection
probabilities by moving more AGN out of the stellar locus than into it.

The eleven objects missed by the revised criteria are noted in
Table~\ref{tab_phot}.
The objects include all the \nev\ selected AGN except 0223.130940,
	which are all morphologically classified galaxies 
	whose colors could be affected by host galaxy emission;
the probable $z=3.08$ object 1447.051899, which has an unusual spectrum and
	unusual colors; 
the $z=3.17$ object 1447.020157, in the redshift range where color selection
	of quasars is difficult because of contamination by F stars;
and two unresolved $z\sim0.65$ objects with the two highest equivalent width
\MgII\ lines in our sample (1447.170990 and 2148.102026), the latter of which
is also a radio-loud AGN.
The $U-B$ colors of these two objects are unusually red for their 
observed redshifts, presumably due to strong \MgII\ emission in the 
observed $B$-band.  However, their $B-V$ colors are not unusually blue,
so the objects were not selected in the $U-B$/$B-V$ diagram.  These quasars'
spectra (which cover the $B$ and $V$ bands) do seem to show somewhat redder
continua than typical quasars.

It is reassuring that color-selected surveys can recover $\sim$81$\pm$6\%
of the CNOC2 sample of non-color-selected AGN.  
This is consistent with the estimates of $\gtrsim$81\% completeness for the
quasar sample of \citet{zit92} by \citet{czm93} and 77$\pm$11\% completeness
for the \citet{kk88} quasar sample by \citet{maj93},
both from proper motion surveys in the fields of earlier color-selected surveys.
Despite the relatively high completeness, it is of some concern that several of
the most unusual CNOC2 AGN would be missed by the color selection discussed
above.
This includes most of the \nev-selected AGN, which have redder colors than
other AGN on average.  This is probably due to greater host galaxy contamination
since they have lower average redshifts and luminosities than the other AGN.
However, it should be possible to increase the 81$\pm$6\% completeness 
even further by selecting blue AGN candidates not just from the $U-B$/$B-V$
color-color diagram, since several 
of the missed objects stand out
in at least one other diagram (Figures~\ref{f_uvvr}--\ref{f_brri}),
by selecting objects blueward of the stellar locus in the bluer color and
redward of it in the redder color 
(e.g., to select the \nev\ AGN in Figure~\ref{f_brri}),
or by searching multidimensional color space directly rather than in 
two-dimensional projections.  

We have neglected the 
morphological selection also used in the DMS to select quasar candidates
(an object had to
be classified unresolved in at least half the filters in which it was detected).
However, resolved AGN account for only $\sim$20\% of all AGN to $R=22.09$ 
(\S\ref{Density}).  Thus AGN samples which exclude resolved objects at worst
miss $\sim$20\% of AGN.\footnote{This estimate is for AGN above a minimum
luminosity of $M_B=-18.85$, the luminosity of our faintest \nev\ AGN.
Fainter AGN are not detectable against the background of their host galaxies
in the CNOC2 spectra.}
However, the actual situation is likely to be better.
The DMS morphological selection criteria were designed to be lenient in their 
exclusion of galaxies, so that 33\% of the $R<23$ unresolved objects in the 
DMS catalog are expected to be galaxies \citep[][Table 3]{hal96dms1}.
CNOC2 has better star-galaxy separation than this, and would probably
resolve many objects unresolved by the DMS and other surveys,
heightening those surveys' apparent incompleteness to resolved AGN.

Determining the fraction of non-AGN in the sample of AGN candidates which 
meet our color selection criteria is beyond the scope of this paper.  
However, we note that 20 of the 65 spectroscopically confirmed galaxies
morphologically misclassified as unresolved would be selected as AGN
candidates by our color selection criteria.  Thus our color selection
criteria do suffer some contamination by compact blue galaxies.

\subsection{Broad Absorption Line AGN (BALs)} 	\label{BALs}

The canonical BAL fraction in quasar surveys is 10\%.  
The fraction in the CNOC2 AGN is highly uncertain but consistent with this,
given the small number statistics involved and the fact that most of our
objects are at low redshift where only the rare LoBALs might be detectable.
Our sample includes only one confirmed BAL (the LoBAL 0223.130940),
but there is also a potential mini-BAL which appears to show narrow absorption
from \CII\ and Si\,{\sc iv}\,$\lambda$1400 but not \CIV\ (0223.051932),
one object with a deep absorption trough and unusual emission-line ratios
which might be a BAL (1447.051899), and several candidate 
Iron LoBALs 
among the absorption-line AGN candidates: 
2148.050488, which resembles FIRST J142703.6+270940, 
and 0223.071845, which resembles FIRST J084044.4+363328 \citep{bea00}.

\subsection{Associated Absorption Lines}	\label{assabs}

From six to nine 
of the 47 emission-line AGN in this sample show strong
associated absorption within $\pm$5000~\kms\ of the emission line redshift
(3--7 in \MgII\ and 2--4 in \CIV, each number being independent except for one
one-line object, 0223.021739, which is assumed to be \mgii\ but could be \civ).
Seven or eight objects have spectra that cover \CIV, for an associated
\CIV\ absorption fraction of between 
29$^{+26}_{-18}$\% and 50$\pm$22\% (1$\sigma$).  
Small number statistics prevent any firm conclusion, but it is still 
instructive to compare this fraction to that observed in other quasar samples.
\citet{fol88} found strong (rest-frame equivalent width
\ew$>$1.5\,\AA) associated absorption in 22 of 88 radio-loud quasars (25\%),
but in only 1 of 29 radio-quiet quasars (3.4\%), all at $z>1.25$.
However, at $z<1.1$, \citet{gea99} found only one strong associated \CIV\ 
absorption system in a sample of 64 $z<1.1$ quasars (25 RQQs and 39 RLQs),
possibly due to redshift evolution.
Given the low luminosity of the CNOC2 AGN compared to the quasars in both
those studies, our observations
may be evidence in favor of the hypothesis of \citet{mjp94} 
that associated \civ\ absorption is more common among
less optically luminous quasars regardless of radio power
\citep[but for a dissenting opinion see][]{fol88}.
However, \citet{ulr88} found strong associated \civ\ or \mgii\ absorption among
only 2.5$-$9\% of $z\lesssim0.1$ Seyfert~1 galaxies 
with $M_B$ similar to or somewhat fainter than the CNOC2 AGN. 
This argues against a simple anticorrelation of the associated \civ\ absorption
fraction with luminosity, and when compared with the CNOC2 AGN sample
may argue in favor of its evolution with redshift.
As this paper was going to press, we became aware of the preliminary results
of a study of $-$30$<$$M_B$$<$$-$26 AGN in which both RLQs and RQQs are found
to have associated \civ absorption fractions of $\sim$33\% \citep{vh00}.
This is in better agreement with our results than with those of \citet{fol88}.
More data is clearly needed.  

From twelve to twenty-two 
objects have spectra that cover \MgII, depending on the number of
single emission line objects in which the line is \MgII.  
Thus the frequency of associated \MgII\ absorption in the CNOC2 AGN is between
14$^{+12}_{-7}$\% and 58$^{+17}_{-18}$\% (1$\sigma$).  
For comparison,
\citet{abe94} found six associated \mgii\ absorption systems in a sample of
56 quasars (11$^{+6}_{-4}$\%), which was itself an excess at the 97--99\%
confidence level compared to the frequency of intervening absorption systems.
Thus the CNOC2 AGN sample is likely to have a higher incidence of associated 
\mgii\ absorption than other samples.  

If the line of sight required to produce associated low-ionization absorption
is dusty, as has been seen in some other samples of AGN \citep{aeb94,bh95,wh97},
the increased incidence of such absorption in the CNOC2 AGN compared to
color-selected samples might be explained, since dust can 
more easily redden quasars out of the latter.
The $U-B$ and $B-V$ colors of the CNOC2 AGN do extend to redder values than
the DMS AGN (Figure~\ref{f_ubbv}), in support of this suggestion.
However, CNOC2 AGN with known associated absorption are not on average redder
than AGN without such absorption, nor are they preferentially
missed by the color selection criteria discussed in \S\ref{Colors}.
It is possible that associated \mgii\ absorption is more common among
less optically luminous AGN, similar to the suggestion of \citet{mj87} for
associated \civ\ absorption, but the \citet{ulr88} sample argues against
this possibility just as it does for \civ, as discussed above.
Another possible explanation arises from the CNOC2 AGN
sample's lack of bias against
objects with resolved spatial structure such as companions, tidal tails or
large host galaxies.  Such environments may be more likely to have
extensive gas envelopes which show up as associated absorption.
However, the fraction of resolved objects is $\sim$53\% 
among those with associated absorption, consistent with the value of $\sim$60\%
for objects without associated absorption in the $0.6<z<3$ broad-line sample.
Overall, it is not clear why the CNOC2 sample
has a higher incidence of associated absorption than other quasar samples.

\subsection{Detection Frequency and Surface Density of Low-Luminosity AGN} \label{Density}

Broad-line and \nev\ AGN are found in only a small fraction of galaxies.
In the range $0.281 < z < 0.685$, where CNOC2 can detect both normal galaxies
to $\sim$$M^*$+1 (via \OII\ at least) and AGN (via \NeVb\ or other broad lines),
only 0.3$\pm$0.1\% of galaxies have such AGN.  
This is consistent with the results of \citet{sch96a}, who found that
only 0.5$\pm$0.3\% of $z\leq1.35$ CFRS galaxies were broad-line AGN.
However, this is only a lower limit to the AGN fraction among such galaxies
since it includes neither narrow-line AGN nor AGN too faint
relative to their host galaxy for CNOC2 to detect them.
\citet{tre96} found that at least 8\% of $z\leq0.3$ field galaxies in the CFRS
were narrow-line AGN (Seyfert~2s or LINERs), whose identification in CNOC2
is outside the scope of this paper.  

The CNOC2 AGN can be used to estimate the surface density of faint AGN.
We include only the one confirmed absorption-line AGN in these estimates.
First we must weight to account for objects not observed or identified
spectroscopically, using the fraction of AGN among observed objects.
The maximum fraction of AGN should equal the fraction among 
objects with spectra good enough for classification.  This includes spectra
with high SNR but no identification, since in spectra of similar SNR it is
easier to identify AGN than stars and many galaxies.  The minimum fraction
of AGN should equal the fraction among all objects with any spectra.
This assumes that broad lines would have been identifiable in all spectra.

We weight each resolved AGN to account for the galaxies without redshifts
at the same apparent magnitude (within $\pm$0\fm25) in the entire catalog.
We weight the unresolved AGN to account for all unresolved objects in the CNOC2
catalog without redshifts which lie within a fixed apparent magnitude range.
This is because to first approximation,
the unresolved objects with spectra in CNOC2 constitute a fair,
random sample of unresolved objects fainter than the brightest magnitude at
which an unresolved object could be misclassified as resolved, and brighter than
the faintest magnitude at which objects can be identified from CNOC2 spectra.
(In reality, even ``unresolved" AGN are likely to be more extended than stars
  and thus more susceptible to being misclassified as a galaxy and selected
  for spectroscopy.  Our AGN surface density will be a firm upper limit
  since we neglect this effect.)
We also use the average detectability of \nev\ in our \nev\ AGN (71$\pm$20\%;
see \S\ref{nevsample}) to account for similar objects missed due to a low SNR.
Finally, to find the surface density, we divide by the total area of CNOC2
(5433.5\sq\arcmin, or 1.5093\sq\arcdeg).  
We do not account for the $1.24\lesssim z \lesssim1.31$ range where no emission
line falls in our spectral range, so this will be a underestimate.

The surface density of resolved AGN is in the range
46$_{-8}^{+15}$ deg$^{-2}$ to 89$_{-16}^{+29}$ deg$^{-2}$
from 31 objects to $R=22.09$, the magnitude of the faintest resolved AGN.
We estimate the bright magnitude limit for unresolved AGN to be $R$=19.5,
since the magnitude distribution of unresolved objects with spectra
shows a dramatic increase faintwards of $R$=19.5
and the brightest unresolved AGN has $R$=19.75.
For ease of comparison with resolved AGN, we use a faint magnitude limit of
$R=22.09$.
This is reasonable since 
the magnitude distribution of unresolved objects with good spectra peaks at
$R$=21 and includes only a few objects with $R>22.09$.
There are 5949 unresolved objects in this magnitude range, of which
346 have spectra, 240 have spectra good enough for classification,
and 17 are AGN (15 automatically classified as stars by PPP,
  plus 2 definitely misclassified as galaxies due to close neighbors).
With additional weighting for \nev\ detectability as discussed above,
and including the $\sim$23 unresolved AGN deg$^{-2}$ with $R<19.5$ \citep{hs90},
we obtain a surface density in the range
221$_{-48}^{+61}$ to 309$_{-69}^{+87}$ unresolved AGN deg$^{-2}$,
for a total surface density in the range
267$_{-49}^{+63}$ to 398$_{-73}^{+92}$ AGN deg$^{-2}$ to $R=22.09$.

For comparison with estimates in the literature, we also calculated our AGN
surface density in the $B$-band 
using $B$ magnitude weights
and adopting a bright magnitude limit of $B=20.2$ for unresolved objects.
We find a total AGN surface density of
244$_{-50}^{+65}$ deg$^{-2}$ to 314$_{-67}^{+87}$ deg$^{-2}$ to $B=22.6$,
compared to 233 deg$^{-2}$ estimated to the same magnitude 
by \citet{maj93} using a 0.3~deg$^2$ color-selected survey augmented with
proper motion and variability studies.
We find a surface density of
197$_{-38}^{+51}$ deg$^{-2}$ to 269$_{-54}^{+73}$ deg$^{-2}$ 
for all $z\leq2.1$ AGN to $B=23.5$, 
compared to 120$_{-70}^{+120}$ deg$^{-2}$ to 210$_{-90}^{+140}$ deg$^{-2}$ 
(depending on the color cut used to reject potential compact galaxies)
estimated by \citet{bwa99} from a 0.025~deg$^2$ morphological- and 
color-selected {\em HST} survey.
Our AGN surface density estimates are comparable to both these previous
estimates, though all of them are still quite uncertain.

\subsection{Equivalent Widths and the Baldwin Effect} 	\label{EWs}

In this section we discuss the emission-line rest-frame equivalent widths (\ew)
of the CNOC2 AGN compared to expectations based on previous studies of AGN.

The \ew\ values and associated uncertainties for all lines
(Tables \ref{tab_spec_bel}-\ref{tab_spec_abs})
were determined using the procedure of \citet{gre96},
except that we neglect the random errors in our continuum fits,
and using the wavelength ranges defined by
\citet{fra91} for the composite LBQS spectrum.
For each line, the continuum was modelled as a single cubic spline fit
to the entire spectrum outside that line's wavelength range,
after a maximum of 10 iterations of $\pm$2$\sigma$\ pixel rejection.  
The continuum determined in this manner was consistent within the uncertainties
with the continuum determined from specific rest-frame regions around each line.
(For AGN with \lya\ emission, only wavelengths redward of the line were
used in the fit, since \lya\ forest absorption depresses the continuum
blueward of the line.  However, we do not correct for absorption atop the
\lya\ profile since such corrections would be highly uncertain.)
Equivalent widths for features in absorption-line objects were found using the
same procedure on wavelength windows extending 25\,\AA\ to either side of the
line(s) being fitted in the rest frame of the AGN.  

\begin{figure}
\epsscale{0.55}
\plotone{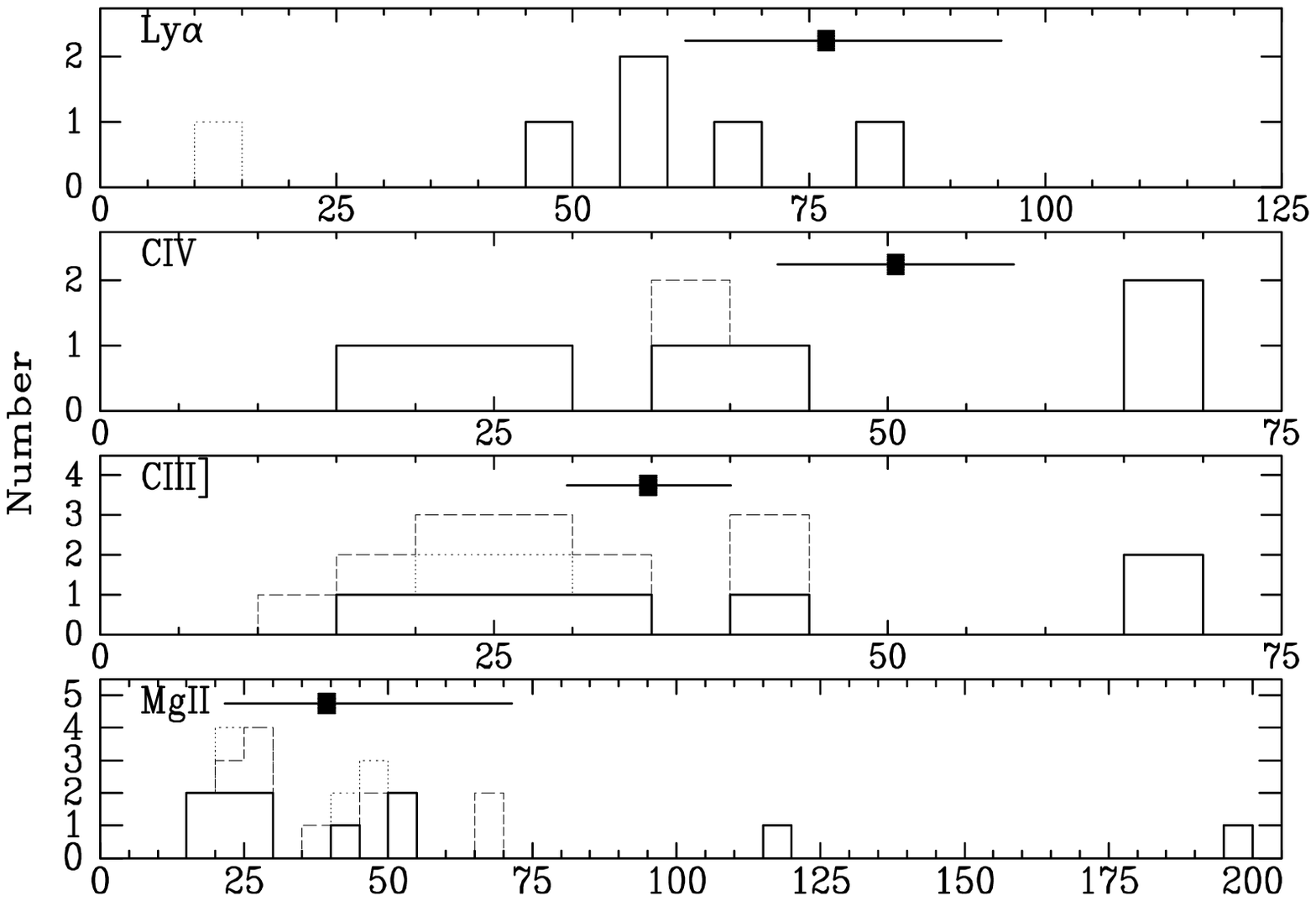} 
\caption[]{
\singlespace
Histograms of measured rest-frame equivalent widths \ew, in \AA,
for CNOC2 broad emission line selected AGN for (from top to bottom) 
\LyA\ (including \NV), 
\CIV, \CIII\ (including \AlIII\ and \SiIII), and \MgII.
Solid histograms are for lines confirmed to be the stated transition; 
dashed histograms for one-line objects where the line was assumed to be
\mgii\ but could be \ciii\ (or \civ\ in one case), and dotted histograms for 
one-line objects where the line was assumed to be \ciii\ or \lya\ but could be \mgii.
The solid points and error bars show the mean and $\pm$1$\sigma$ RMS \ew\ for
the average luminosity of the quasars (see \S\ref{EWs}), as predicted by
\citet{zam92} for \mgii\ and \citet{gre96} for \ciii, \civ\ and \lya.
}\label{f_ews}
\end{figure}

Figure~\ref{f_ews} shows histograms of \ew\ for \lya\ (including \nv), \civ,
\ciii\ (including \aliii\ and \siiii) and \mgii\ for the CNOC2 broad emission
line selected AGN.  
Solid histograms are for lines confirmed to be the stated transition;
dashed histograms for one-line objects where the line was assumed to be
\mgii\ but could be \ciii\ (or \civ\ in one case), and dotted histograms for
objects where the line was assumed to be \ciii\ or \lya\ but could be \mgii.
Our \ew\ measurements generally fall within the range seen in the literature
\citep{zam92,hs90}.  Even the two objects with very high \ew(\mgii) emission lie
within the range of previously observed values.
However, our \ew\ distributions tend to be skewed lower than the average values
predicted by previous studies of the Baldwin effect \citep[the anticorrelation
of \ew\ and luminosity; for a recent review see][]{os99}.
The solid points and error bars in Figure~\ref{f_ews} show the mean \ew\ and
$\pm$1$\sigma$ standard deviation 
in the \ew\ predicted for that line at the average
luminosity of the quasars in each histogram.  These predictions are taken
from \cite{zam92} for \mgii\ and \cite{gre96} for \ciii, \civ\ and \lya.
In the following sections we discuss each of
the transitions in Figure~\ref{f_ews} in turn.  

\subsubsection{The Baldwin Effect in \MgII}  \label{bemgii}	

We compare \ew(\mgii) for our $0.6<z<1.2$ AGN
with the $z<1.5$ sample of \citet[][hereafter Z92]{zam92}.
We compute the monochromatic magnitudes $M_{2798}$ for our objects following
Eq.~1 of Z92: $M_{\lambda}=M_B-2.5~\alpha~{\rm log}(\lambda/4400)$, with
continuum power-law slope $\alpha=0.5$.  However, our $M_B$ were calculated 
using $q_0=0.5$ instead of $q_0=1$ and $k$-corrections appropriate for
$\alpha=0.7$ \citep{cv90} instead of $\alpha=0.5$.
The former results in negligible differences, while the latter
biases our magnitudes fainter than those of Z92 by $\lesssim$0\fm15,
again negligible since the Baldwin effect is not steeply magnitude dependent.

We identify nineteen objects with \mgii\ emission, though the true number may
be anywhere in the range 12 to 22 (some lines identified as \MgII\ may turn out
  to be \CIII, and vice versa).  These nineteen have average 
$M_{2798} = -21.5\pm0.2$, 
and average \ew(\mgii)=49$\pm$10\,\AA\ (in all our \ew\ discussions 
  our quoted uncertainties are $1\sigma$ standard deviations of the mean).
The \ew(\mgii) predicted by Z92 at this $M_{2798}$ is
38$_{-5}^{+11}$\,\AA.  
Even in the extreme cases of 12 or 22 \mgii\ identifications,
our average \ew(\mgii) agrees with the Z92 prediction to within $\simeq1\sigma$.
Thus the \mgii\ Baldwin effect in the $0.6<z<1.2$ CNOC2 AGN 
is consistent with the \mgii\ Baldwin effect measured by Z92 for $z<1.5$ AGN.

\subsubsection{The Baldwin Effect in \CIII}  \label{beciii}

We compare \ew(\ciii+\aliii+\siiii), hereafter simply \ew(\ciii),
for our $1.35<z<2.2$ AGN with the sample of \citet[][hereafter G96]{gre96},
in which \ew(\ciii) are available only for $z<0.7$ objects.

For this comparison we require the monochromatic luminosity density
at rest-frame 1450\,\AA, $L_{\nu}$(1450\,\AA), called $l_{UV}$ by G96.
We calculate $L_{\nu}$(1450\,\AA) for the CNOC2 AGN from our broadband
photometry even when our spectra cover 1450\,\AA\ in the rest frame because
our observing procedures are not designed to yield accurate spectrophotometry.
We convert the $U$ and $B$ magnitudes to monochromatic flux densities at their
effective wavelengths \citep{fsi95} and interpolate to find $f_{\nu}$ at rest
wavelength 1450\,\AA, from which we calculate $L_{\nu}$(1450\,\AA).

\begin{figure}
\epsscale{0.6}
\plotone{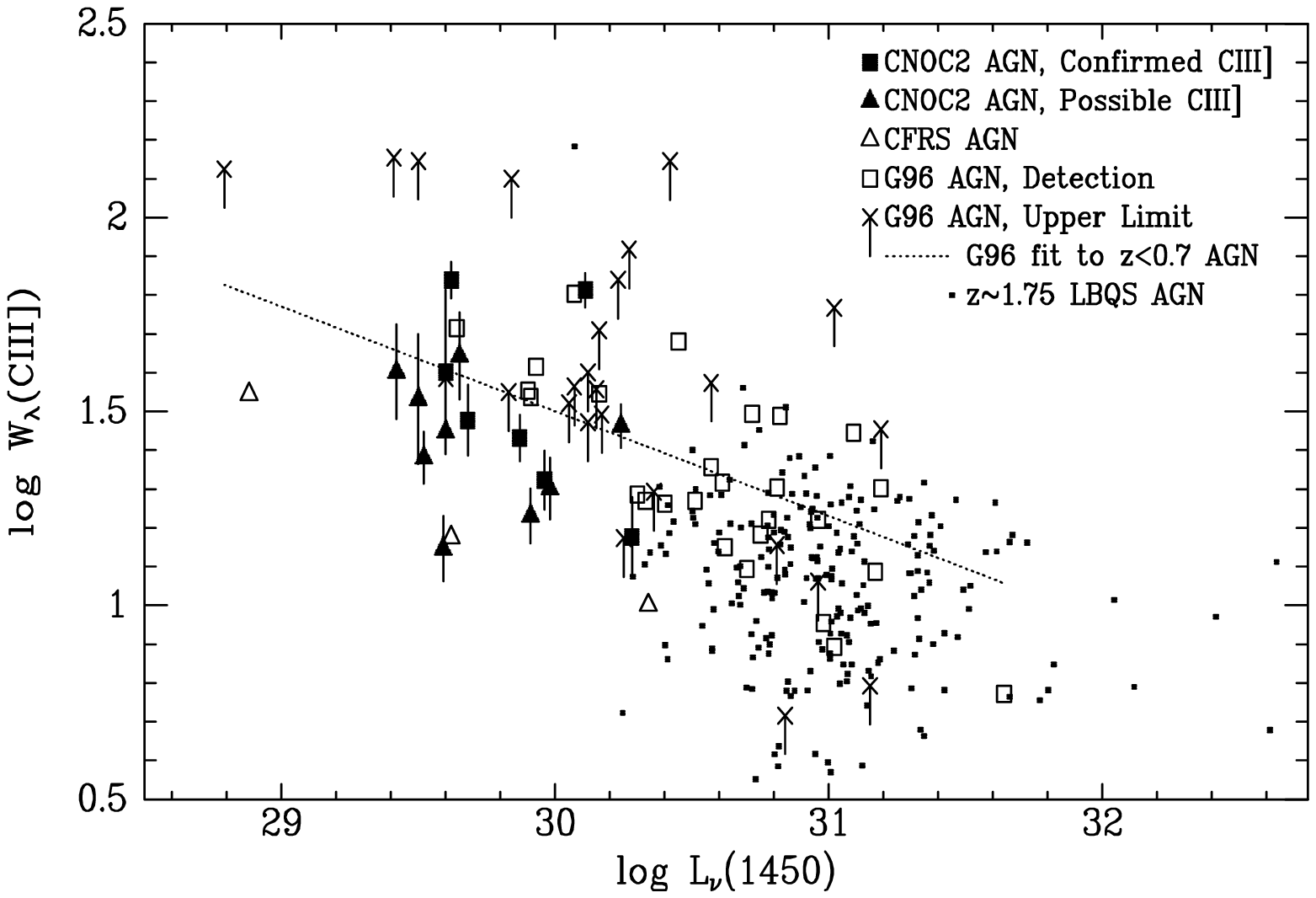} 
\caption[]{
\singlespace
The Baldwin effect for \ciii.
Filled squares are CNOC2 AGN with identified \ciii, 
filled triangles are CNOC2 AGN with possible \ciii,
open triangles are AGN from the CFRS (see text),
small dark points are AGN from the LBQS with $z\sim1.75$,
open squares are AGN from \citet{gre96} with detected \ciii\ 
and crosses with error bars are AGN from \citet{gre96} with only upper limits.
The dotted line shows the correlation found by \citet{gre96}
over the luminosity range of his sample.
}\label{f_beciii}
\end{figure}

We identify nine objects with \ciii\ emission
(though the true number may be anywhere in the range 7--16), with average 
$L_{\nu}$(1450\,\AA)$=8.4\pm0.6~10^{29}$ ergs~s$^{-1}$~Hz$^{-1}$
and average \ew(\ciii)=35$\pm$6\,\AA.  
The \ew(\ciii) predicted by G96 at this $L_{\nu}$(1450\,\AA) is
35$\pm$2\,\AA.  
Thus our average \ew(\ciii) agrees with the prediction of G96,
as illustrated in Figures~\ref{f_ews} and \ref{f_beciii}.
This agreement would remain even if all possible \ciii\ identifications
turned out to be \ciii.
However, Figure~\ref{f_beciii} shows that the majority of the CNOC2 AGN would
lie below the \ew(\ciii) relation found by G96 in that case.
Also lying predominantly below the relation are three $1<z<2$ AGN from the
Canada-France Redshift Survey \citep[CFRS; ][open triangles]{sch96a} and
$\sim$250 $1.35<z<2.2$ AGN from the LBQS \citep[][small dark points]{fra92}.
The combined CNOC2, LBQS and CFRS data appear to have a \ciii\ Baldwin effect
with a somewhat lower normalization than the G96 sample.  The offset is not
formally statistically significant, however, and is small enough that it could
conceivably be caused by systematic differences in \ew\ measurement procedures.

\subsubsection{The Baldwin Effect in \CIV}  \label{beciv}

\begin{figure}
\epsscale{0.65}
\plotone{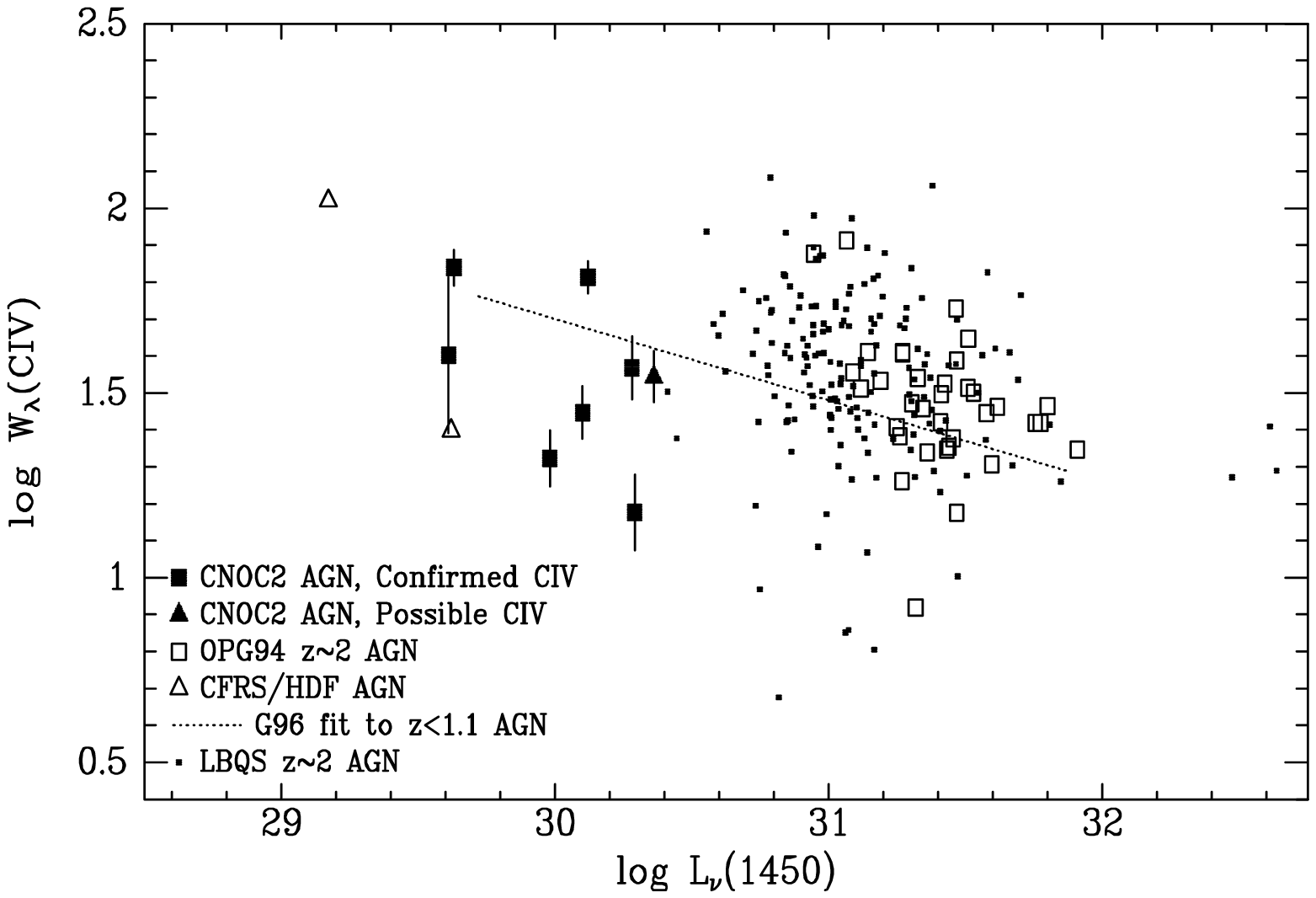} 
\caption[]{
\singlespace
The Baldwin effect for \civ.
Filled squares are CNOC2 AGN with identified \civ,
filled triangles are CNOC2 AGN with possible \civ,
small dark points are $z\sim2$ AGN from the LBQS,
open squares are $z\sim2$ AGN from \citet{opg94},
and the open triangles are AGN from the CFRS and HDF surveys. 
The dotted line shows the correlation found by \citet{gre96} for $z<1.6$ AGN
the luminosity range of his sample.
Some LBQS AGN with very low \ew(\civ) values may be affected by BAL absorption.
}\label{f_beciv}
\end{figure}

We first compare \ew(\civ) for our $1.9<z<2.55$ AGN 
with the \civ\ Baldwin effect measured at $z<1.1$
by G96 using $L_{\nu}$(1450\,\AA) (see \S\ref{beciii}).
There are seven objects with \civ\ in the CNOC2 spectra
(and possibly an eighth, 0223.021739), with average
$L_{\nu}$(1450\,\AA)$=1.2\pm0.2~10^{30}$ ergs~s$^{-1}$~Hz$^{-1}$
and average \ew(\civ)=33$\pm$6\,\AA.  
The \ew(\civ) predicted by G96 at this $L_{\nu}$(1450\,\AA) is
51$^{+5}_{-4}$\,\AA.  
Thus our average \ew(\civ) is lower than the prediction of G96 at $2.5\sigma$
significance, as illustrated in Figure~\ref{f_ews} and Figure~\ref{f_beciv}.

However, the G96 sample measures \civ\ only at $z<1.1$,
whereas the CNOC2 AGN with measured \civ\ have $1.9<z<2.55$.
To examine the possibility of redshift dependence, in Figure~\ref{f_beciv}
we compare to the $1.8<z<2.2$ sample of \citet[][hereafter OPG94]{opg94}
and to LBQS AGN with $1.8<z<2.55$ \citep{fra92}.
Together these samples show evidence for a shallower 
\civ\ Baldwin effect at $1.8<z<2.55$ than at $z<1.1$.
Some of the LBQS AGN with the smallest \ew(\civ) are likely affected by
broad absorption lines, but this would not affect the above conclusion.
The open triangles in Figure~\ref{f_beciv} show that the lower than expected
\ew(\civ) values for low-luminosity AGN found in CNOC2 are supported by
measurements of one $z\simeq2$ AGN from the CFRS \citep{sch96a}, but that
a $z\simeq3.4$ AGN from the HDF \citep{bea99,coh00} shows a high \ew(\civ).  

\subsubsection{The Baldwin Effect in \LyA+\NV}  \label{belya}

We first compare \ew(\lya+\nv), hereafter simply \ew(\lya),
for our 3$<$$z$$<$4 AGN with the \lya\ Baldwin effect measured at $z<1.6$
by G96 using $L_{\nu}$(1450\,\AA) (see \S\ref{beciii}).
There are five objects with \lya\ in the CNOC2 spectra
(and possibly a sixth, 1447.051899), with average
$L_{\nu}$(1450\,\AA)$=1.6\pm0.4~10^{30}$ ergs~s$^{-1}$~Hz$^{-1}$
and average \ew(\lya)=63$\pm$6\,\AA.  
The \ew(\lya) predicted by G96 at this $L_{\nu}$(1450\,\AA) is
77$^{+8}_{-7}$\,\AA.  
Four of our five objects have \ew(\lya) below this predicted average
(on average they are a factor of 1.3$\pm$0.1 low), although our average
\ew(\lya) is consistent with the predictions at the $\simeq1.5\sigma$ level
due to the large uncertainties.

However, the G96 sample measures \lya\ only at $z<1.6$,
whereas the CNOC2 AGN with measured \lya\ have $3<z<4$.
To examine the possibility of redshift dependence, we compare the CNOC2 AGN
to the $z\sim3$ sample of OPG94 in Figure~\ref{f_belya}
and find no evidence for a Baldwin effect in \lya\ at $z\simeq3-4$.  The typical
\ew(\lya) is instead roughly constant from the brightest LBQS AGN down to a very
faint $z\simeq3.4$ AGN from the HDF \citep[][open triangle]{bea99,coh00}.
This stands in contrast to 
the $z<1.6$ sample of G96 and the $z<1$ IUE sample of OPG94.
However, \citet{ssg91} noted that the \ew(\civ) of their sample
of $z>4$ quasars with $L_{\nu}$(1450)$\sim$10$^{30.5-31}$
fell below the predicted Baldwin effect by a factor $\sim1.4$.
Moreover, a comparison of composite $z\gtrsim2$ spectra from \citet{boy90} and
the LBQS 
also showed
no appreciable \lya\ Baldwin effect \citep[their Figures 5 and 6]{fk95}.

\subsubsection{The Baldwin Effect:  Discussion}  \label{becnoc2}

Broad-line AGN have quite similar spectra over a range of $\sim$10$^6$ in
luminosity, implying that the broad-line region parameters which determine
the emergent spectra scale nearly uniformly with luminosity.  
A Baldwin effect for a given line means that this scaling
is not exactly homologous for that line.
The lower than expected average \ew\ of CNOC2 AGN broad emission lines suggests 
that the normalization of the Baldwin effect may be lower in \ciii\ than in
some previous studies, and that the slope of the Baldwin effect in \lya\ and
\civ\ may evolve with redshift, steepening with cosmic time.
These differences are too large to be explained by systematic errors
in our absolute magnitude or luminosity determinations,
due to the relatively shallow slope of the Baldwin effect.
Variability effects should also be negligible since most images and spectra
in a given CNOC2 field were taken within a few days of each other.  
Associated absorption may be present in up to $\sim$50\% of
the \mgii\ and \civ\ lines in these AGN (\S\ref{assabs}), but it is not strong
enough to significantly affect the \ew\ distributions of these lines.
The strongest such absorption, in 0920.092046,
does reduce its \ew(\civ) by almost 50\%,
but the average \ew\ reduction due to associated absorption is $\lesssim$10\%
in \civ\ and $\lesssim$5\% in \mgii.
For \lya, we do not correct for associated or \lya\ forest absorption atop the
emission line, but neither do OPG94; so, this should not bias our comparison.

\begin{figure}
\epsscale{0.7}
\plotone{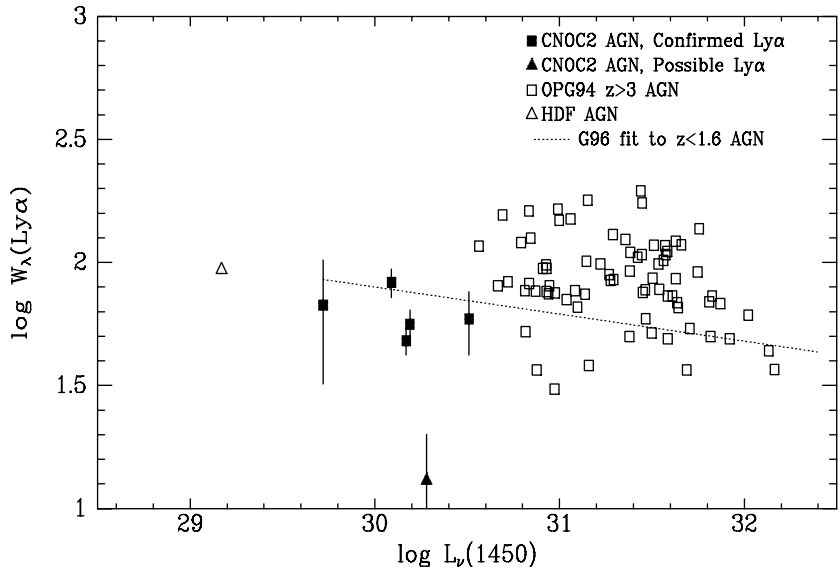} 
\caption[]{
\singlespace
The Baldwin effect for \lya.
Filled squares are CNOC2 AGN with identified \lya,
filled triangles are CNOC2 AGN with possible \lya,
open squares are $z>3$ AGN from \citet{opg94},
and the open triangle is an AGN from the HDF (see text).
The dotted line shows the correlation found by \citet{gre96} for $z<1.6$ AGN
over the luminosity range of his sample.
}\label{f_belya}
\end{figure}

Probably the most important caveat is the unknown effect of 
the different selection criteria and \ew\ measurement techniques
of the various samples on the measured Baldwin effect.
For example, the OPG94 $z>3$ sample consists mainly of objects discovered
via slitless spectroscopy, and thus might be biased against weak-lined AGN.
The OPG94 $z>3$ AGN do have higher \ew(\lya) at $L_{\nu}$(1450\,\AA)=10$^{31-32}$
than the G96 $z<1.6$ AGN (Figure~\ref{f_belya}), which could arise from this
possible bias or from different \ew\ measurement techniques.
In principle, CNOC2 might be more sensitive to weak-lined AGN that other
surveys due to its lack of strong selection criteria.  If true, this would mean
that other surveys overestimate the average \ew\ at all luminosities, and our
lower average \ew\ are simply a better estimate of the true average \ew.
However, Figs.~\ref{f_beciii}-\ref{f_beciv} show that the G96, LBQS, and OPG94
samples all include objects with \ciii\ or \civ\ lines as weak as in CNOC2
(the same is true for the Z92 sample and \mgii).
The detection of some weak-lined objects in these samples
argues against their having a strong bias against such objects.
However, Fig.~\ref{f_belya} suggests that such a bias may exist
for \lya\ in the OPG94 sample.

Another worry is whether other parameters might systematically
bias the observed Baldwin relation in a given sample.
\citet{wea99} see a \civ\ Baldwin effect in their $z<1$ sample only if they
exclude narrow-line Seyfert~1 objects, which have narrow 
H$\beta$ lines and systematically weak \ew(\civ),
effectively flattening the slope at low luminosity.  
If any of the CNOC2 AGN fall into this category, it 
might explain their low \ew(\civ) but not their low \ew(\lya).

Redshift evolution of the Baldwin effect has rarely been discussed in the
literature, since very few line measurements for low-luminosity AGN at
$z\gtrsim1$ are available.  
At the low luminosities of the CNOC2 AGN, the \lya\ Baldwin effect has only 
been previously studied at $z\lesssim0.5$ \citep{opg94,gre96,wea99}.  
The CNOC2 AGN provide some of the first evidence that the slope of the Baldwin
effect may evolve to steeper values with cosmic time in \lya\ and \civ,
although some evidence of this effect is present in
\citet{ssg91} and \citet{fk95}.

A currently favored model explains the Baldwin effect as a result of the
luminosity dependence of AGN continuum spectral energy distributions or SEDs
\citep{ms82,os99}, wherein more luminous AGN have softer ionizing continua.
Evolution in the Baldwin effect might indicate evolution with cosmic time
of AGN SEDs at a given luminosity, which might arise from evolution in the
relation between AGN luminosity and black hole mass \citep{wan99a}.
However, an earlier model for the \civ\ Baldwin effect which postulated a 
smaller ionization parameter $U$ (ratio of ionizing photon to gas densities)
in more luminous AGN predicted no \lya\ Baldwin effect, or even an 
anti-Baldwin effect \citep{mf84,sf93}.
An anticorrelation between $U$ and $L$ has also
been recently suggested on entirely independent grounds \citep{kea00}.
Given the theoretical implications, further data on high-$z$ low-luminosity
AGN from well-understood samples is needed to confirm our possible evidence for
redshift evolution of the slope of the Baldwin effect in \lya\ and \civ.

\section{Conclusions}	\label{Conclusions}

We have presented a sample of 47 
confirmed and 14 
candidate AGN from the CNOC2 field galaxy redshift survey,  
with redshifts from $z=0.2697$ to $z=4.6755$.  
In this paper we have discussed the following:

1. The average absolute magnitude of these AGN is $M_B \simeq -22.25$,
below the quasar/Seyfert division at $M_B = -23$ (\S\ref{zMB}).
Only two AGN have been detected as radio sources;
both of those are radio-loud (\S\ref{magscoordsids}).

2. Spectroscopy was
preferentially obtained for objects morphologically classified as resolved,
but several hundred unresolved objects were also observed inadvertently or
serendipitously, yielding samples of unresolved as well as resolved AGN.
The unresolved and resolved AGN are expected to be representative 
of their particular types, but as a whole the sample may be biased toward
host galaxies of above average luminosity relative to the AGN luminosity
(\S\ref{morphs}).

3. No color selection criteria were involved in selecting this sample.
Simple color-color diagram selection criteria can recover $\sim$81$\pm$6\%
of the CNOC2 AGN.  However, several of the most unusual objects would still be
missing from such a color-selected sample.  
It should be possible to increase the completeness of color-selected samples 
even further by selecting AGN candidates in multidimensional color space
(\S\ref{Colors}).

4. The emission line selected AGN subsample shows
a higher incidence of associated \MgII\ absorption than in previous surveys and
an incidence of associated \CIV\ absorption which may be more similar to that of
radio-selected quasar samples than optically-selected ones (\S\ref{assabs}).
The sample size is very small (at most 9 objects with associated absorption),
but this may represent 
an anti-correlation of strong associated absorption with optical luminosity 
(although any such trend does not appear to continue in nearby Seyfert~1
  galaxies of even lower luminosity),
or a decreasing frequency of its occurrence with cosmic time.
Alternatively, it may arise either because the selection method is not biased
against objects with resolved spatial structure 
(if such luminous and/or interacting galaxies preferentially stir up the gas
  responsible for associated absorption),
nor against redder objects 
(if associated absorbers are dusty).
The admittedly small CNOC2 AGN sample provides no strong evidence 
to support either of these last two possibilities, however.

5. We find that only 0.3$\pm$0.1\% of galaxies brighter than $\sim$$M^*$+1
at $0.281 < z < 0.685$ contain broad-line or \nev\ AGN.
We find a total surface density in the range
267$_{-49}^{+63}$ to 398$_{-73}^{+92}$ AGN deg$^{-2}$ to $R=22.09$.
This AGN surface
density is comparable to previously published estimates (\S\ref{Density}).
About 20\% of these AGN are classified as resolved or probably resolved in CFHT
seeing and might be missed in surveys which target unresolved objects only.

6. The average rest-frame equivalent widths for \lya\ and \civ\ are less than
predicted by previous studies of the Baldwin effect, while the equivalent widths
for \ciii\ and \mgii\ are in agreement with previous work.
The data suggest that slope of the Baldwin effect in \lya\ and \civ\ evolves
with redshift, steepening with cosmic time.  
If confirmed, this may 
indicate evolution of AGN SEDs or ionization parameters with $z$ (\S\ref{EWs}).

7. The sample includes several unusual objects:
one with a candidate double-peaked \mgii\ emission line of very high equivalent width
	(2148.102028; \S\ref{belnotes}),
several with unusual emission line properties 
	(0223.130495 and 1447.051899; \S\ref{belnotes}),
one with an \oiii\,$\lambda$3133 broad absorption line,
	(0223.130940; \S\ref{nevnotes})
and at least one with an optical absorption-line spectrum but broad H$\alpha$
	emission in the near-IR (2148.050488; \S\ref{abssample}).

The CNOC2 AGN form a unique spectroscopically-selected sample useful for
studying the properties of low-luminosity AGN, as this paper has shown
in their possible deviation from the expected Baldwin effect.
Planned followup observations include
further spectroscopy to study the Baldwin effect
and associated absorption in as many objects as possible,
to confirm absorption-line AGN candidates
	and the unusual properties of certain objects,
and to resolve the \mgii/\ciii\ redshift degeneracy where present.

\acknowledgements
\noindent{We thank CTAC and the CFHT for generous allocations of telescope time,
the CFHT operators for their expert and dedicated assistance during observing,
D. Balam for assistance with the astrometry,
P. Francis for providing LBQS data,
J. Cohen for sharing her faint AGN spectra,
E. Hooper for providing $k$-corrections electronically,
and C. Steidel for providing his spectrum of cB58.
CNOC was supported by a Collaborative Program grant from NSERC,
as well as by individual NSERC operating grants to HY and RC.
HL acknowledges support provided by NASA through Hubble Fellowship grant
\#HF-01110.01-98A awarded by the Space Telescope Science Institute, which
is operated by the Association of Universities for Research in Astronomy,
Inc., for NASA under contract NAS 5-26555.
}

\footnotesize

\begin{deluxetable}{ccrccccccc} 
\tablecaption{Spectroscopic Data for CNOC2 Broad Emission Line AGN\label{tab_spec_bel}}
\rotate
\tabletypesize{\small}
\tablewidth{635.00000pt}
\tablenum{1}
\tablehead{
\colhead{}   & \colhead{}    & \colhead{}    & \colhead{log $L_{\nu}$}    & \colhead{Ly$\alpha$/\nv} & \colhead{\sio} & \colhead{\civ} & \colhead{\ciii} & \colhead{\mgii} & \colhead{}      \\[.2ex]
\colhead{ID} & \colhead{$z$} & \colhead{$R_{cor}$} & \colhead{(1450\,\AA)} & \colhead{1216/1240}           & \colhead{1400} & \colhead{1549} & \colhead{1909}  & \colhead{2798}  & \colhead{Other}
}
\startdata
0223.190359 & 0.26970$\pm$0.00030 & 7.92 & \nodata & \nodata & \nodata & \nodata & \nodata     & \nodata   & \OII, 8$\pm$2 \\
            &                     &      &   	   &         &         &         &             &           & H$\gamma$, 5$\pm$3: \\
            &                     &      &   	   &         &         &         &             &           & \HeIIfs, 7$\pm$3: \\
            &                     &      &   	   &         &         &         &             &           & H$\beta$, 4$\pm$3 \\
0223.130495 & 0.52811$\pm$0.00038 & 5.46 & \nodata & \nodata & \nodata & \nodata & \nodata     & \nodata   & \OII, 3.4$\pm$0.3 \\
2148.151660 & 0.60233$\pm$0.00038 & 3.00 & \nodata & \nodata & \nodata & \nodata & \nodata     & 26$\pm$4: & \OIII\,$\lambda$3133, 8$\pm$2? \\
            &                     &      &   	   &         &         &         &             &           & \OII, 1.4$\pm$1.3 \\
            &                     &      &   	   &         &         &         &             &           & \NeVa, $<$1.5 \\
            &                     &      &   	   &         &         &         &             &           & \NeVb, 3$\pm$2? \\
0920.081010 & 0.64744$\pm$0.00036 & 6.24 & \nodata & \nodata & \nodata & \nodata & \nodata     & 16$\pm$3  & \OII, 2.4$\pm$0.6 \\
2148.102026 & 0.65243$\pm$0.00030 & 9.38 & \nodata & \nodata & \nodata & \nodata & \nodata   & 197$\pm$5a? & \NeVa, 5.6$\pm$1.4 \\
            &                     &      &   	   &         &         &         &             &           & \NeVb, 14$\pm$1 \\
            &                     &      &   	   &         &         &         &             &           & \OII, 12$\pm$1 \\
1447.170990 & 0.66707$\pm$0.00075 & 3.64 & \nodata & \nodata & \nodata & \nodata & \nodata     & 22$\pm$11 & \OII, 6$\pm$3 \\
1447.051040 & 0.69900$\pm$0.00300 & 4.00 & \nodata & \nodata & \nodata & \nodata & \nodata    & 54$\pm$19a & \nodata \\		
1447.050689 & 0.70100$\pm$0.00500 & 4.00 & (29.43) & \nodata & \nodata & \nodata & (40$\pm$13) & 27$\pm$13 & \nodata \\		
2148.040045 & 0.72774$\pm$0.00127 & 4.00 & (30.24) & \nodata & \nodata & \nodata & (29$\pm$4)  & 48$\pm$4  & \nodata \\		
2148.031234 & 0.73200$\pm$0.00600 & 4.00 & (29.92) & \nodata & \nodata & \nodata & (17$\pm$3)  & 25$\pm$3  & \nodata \\		
0920.182673 & 0.74002$\pm$0.00092 & 4.00 & \nodata & \nodata & \nodata & \nodata & \nodata     & 21$\pm$4  & \NeVa, 0.5$\pm$1.2 \\ 
            &                     &      &   	   &         &         &         &             &           & \NeVb, 2.8$\pm$1.5 \\
1447.161148 & 0.77000$\pm$0.02000 & 4.00 & (29.59) & \nodata & \nodata & \nodata & (14$\pm$3)  & 21$\pm$3  & \nodata \\		
1447.090674 & 0.79000$\pm$0.03000 & 5.00 & \nodata & \nodata & \nodata & \nodata & \nodata     & 50$\pm$4a & \nodata \\
1447.081984 & 0.84100$\pm$0.00900 & 4.00 & (29.53) & \nodata & \nodata & \nodata & (24$\pm$4)   & 35$\pm$5  & \nodata \\		
0920.080251 & 0.86100$\pm$0.01600 & 4.00 & \nodata & \nodata & \nodata & \nodata & \nodata       & 118$\pm$22a & \nodata \\
0223.021739 & 0.88600$\pm$0.02900 & 5.00 & (30.36) & \nodata & (5$\pm$5:) & (35$\pm$6a) & \nodata & 66$\pm$6a & \nodata \\
2148.110259 & 0.88800$\pm$0.01400 & 4.00 & (29.65) & \nodata & \nodata & \nodata & (44$\pm$13)   & 66$\pm$14 & \nodata \\		
0920.140041 & 0.94000$\pm$0.01700 & 4.00 & (29.50) & \nodata & \nodata & \nodata & (34$\pm$16)   & 47$\pm$17 & \nodata \\		
2148.011928 & 1.06300$\pm$0.01100 & 4.00 & \nodata & \nodata & \nodata & \nodata & \nodata       & 43$\pm$33a? & \nodata \\		
1447.090792 & 1.12000$\pm$0.02000 & 4.00 & \nodata & \nodata & \nodata & \nodata & \nodata       & 28$\pm$3:   & \nodata \\		
1447.150051 & 1.16508$\pm$0.00108 & 5.00 & \nodata & \nodata & \nodata & \nodata & \nodata       & 16$\pm$10   & \nodata \\
\\ \\ \\ \\ \\ \\ \\
2148.060762 & 1.38000$\pm$0.00500 & 4.00 & 29.98   & \nodata & \nodata & \nodata &     20$\pm$4: & (23$\pm$6:) & \nodata \\		
1447.162128 & 1.77000$\pm$0.03000 & 4.00 & 29.60   & \nodata & \nodata & \nodata &     28$\pm$4  & (40$\pm$6) & \nodata \\		
0223.151505 & 1.82835$\pm$0.00087 & 5.00 & 29.88   & \nodata & \nodata & \nodata &     27$\pm$4  & \nodata & \nodata \\		
0920.020246 & 1.94300$\pm$0.01104 & 5.00 & 30.29   & \nodata & \nodata & 15$\pm$4a?: & 10$\pm$3  & \nodata & \HeIIsf, 6$\pm$3 \\
1447.131920 & 1.96500$\pm$0.00300 & 5.00 & 29.63   & \nodata & \nodata &    69$\pm$8 & 38$\pm$8  & \nodata & \HeIIsf, 10$\pm$6 \\
1447.171440 & 2.00000$\pm$0.02000 & 5.00 & 30.12   & \nodata & \nodata &    65$\pm$7 & 44$\pm$7: & \nodata & \nodata \\
1447.132036 & 2.02000$\pm$0.01000 & 5.00 & 29.61   & \nodata & \nodata &  40$\pm$25a & 32$\pm$22 & \nodata & \HeIIsf, 17$\pm$16 \\
0223.030993 & 2.09200$\pm$0.00100 & 5.00 & 29.98   & \nodata & \nodata &    21$\pm$4 & 39$\pm$6 & \nodata & \nodata \\
0920.092046 & 2.45000$\pm$0.01000 & 4.00 & 30.10   & \nodata &    $<$4 &  28$\pm$5a: & \nodata & \nodata & \nodata \\
0223.051932 & 2.51571$\pm$0.04000 & 5.00 & 30.28   & \nodata   & 5$\pm$4a & 37$\pm$8 & \nodata & \nodata & \CII, 3$\pm$2a: \\
1447.051899 & 3.08000$\pm$0.75000 & 3.00 & 30.28   & 13$\pm$7a?: & $<$5    & \nodata & \nodata & (46$\pm$12a?:) & \CII, 4$\pm$3 \\
2148.112371 & 3.15700$\pm$0.00600 & 5.00 & 30.20   & 56$\pm$8    & $<$9    & \nodata & \nodata & \nodata & \nodata \\
1447.020157 & 3.17000$\pm$0.00800 & 5.00 & 30.17   & 48$\pm$6    &    $<$6 & \nodata & \nodata & \nodata & \nodata \\
2148.171412 & 3.19543$\pm$0.00101 & 5.00 & 30.09   & 83$\pm$11   &  $<$6:  & \nodata & \nodata & \nodata & \nodata \\
0223.151510 & 3.44990$\pm$0.01040 & 5.00 & 29.73   & 67$\pm$35:  & $<$47:  & \nodata & \nodata & \nodata & \nodata \\
0223.010767 & 4.00188$\pm$0.00847 & 5.00 & 30.52   & 59$\pm$17:  & \nodata & \nodata & \nodata & \nodata & \lyb/\ovi, 10$\pm$10: \\
0223.011745 & 4.67550$\pm$0.00350 & 5.00 & \nodata & \nodata     & \nodata & \nodata & \nodata & \nodata & \lyb/\ovi, 18$\pm$6: \\
\enddata
\tablecomments{
$R_{cor}$ measures the reliability of the redshift.
Non-integer values denote the SNR of the cross-correlation function peak,
available only for objects with at least one narrow line.
Integer values were assigned by hand, with 5 being a secure redshift and lower
values indicating some uncertainty
(e.g., degeneracy between \MgII\ and \CIII). 
For objects with measured \ciii, \civ\ or \lya, we give the monochromatic
luminosity density at rest-frame 1450\,\AA, $L_{\nu}$(1450\,\AA),
calculated assuming \ho=50~\kmsm, \qo=0.5\ and $\Lambda$=0.  Parentheses
indicate measurements for the alternate redshift discussed in the text.
For equivalent width measurements,
the letter `a' indicates associated absorption 
(a question mark means the reality of the absorption is uncertain),
measurements uncertain for any reason are denoted by colons, and
entries in parentheses are for the alternate redshifts discussed in the text.
\CIII\ measurements include emission from \AlIII\ and \SiIII.
}
\end{deluxetable}

\begin{deluxetable}{ccrrrrrr} 
\tablecaption{Spectroscopic Data for CNOC2 [Ne\,{\sc v}] AGN\label{tab_spec_nev}}
\rotate
\tabletypesize{\small}
\tablenum{2}
\tablehead{
\colhead{}   & \colhead{}    & \colhead{}    & \colhead{\oiii} & \colhead{\nev} & \colhead{\nev} & \colhead{\oii} & \colhead{\neiii/\hei} \\[.2ex]
\colhead{ID} & \colhead{$z$} & \colhead{$R_{cor}$} & \colhead{3133}           & \colhead{3346} & \colhead{3426} & \colhead{3727}  & \colhead{3869/3889}
}
\startdata
2148.170418 & 0.31874$\pm$0.00034 & 5.89 & \nodata     & 9$\pm$5 & 24$\pm$4    & 16$\pm$2    & 18$\pm$2    \\ 
0223.040997 & 0.39868$\pm$0.00030 & 9.07 & \nodata     & $<$3.7  & 12$\pm$4    & 31$\pm$2    & 14$\pm$2    \\ 
1447.150676 & 0.46667$\pm$0.00030 & 8.39 & $<$1.2      & $<$0.9  & 1.3$\pm$0.9 & 6.7$\pm$0.6 & 2.0$\pm$1.4 \\ 
1447.080513 & 0.48287$\pm$0.00044 & 4.03 & $<$1.2      & $<$5.5  & 8$\pm$5     & 11$\pm$3    & 8$\pm$4     \\ 
1447.140343 & 0.50796$\pm$0.00034 & 6.24 & $<$4.2      & 3$\pm$3 & 10$\pm$3    & $<$1.7:     & 4$\pm$3:    \\ 
0223.130940 & 0.53942$\pm$0.00044 & 4.00 & 1.3$\pm$1.2 & $<$0.86 & 5.4$\pm$1.0 & 2.8$\pm$0.8 & 4$\pm$2:    \\ 
0920.161307 & 0.56147$\pm$0.00030 & 9.01 & 4.3$\pm$4.8: & 11$\pm$4 & 23$\pm$4  & 92$\pm$3    & 14$\pm$4:   \\ 
0920.020116 & 0.56225$\pm$0.00040 & 5.00 & 16$\pm$4:   & $<$2.4  & 9$\pm$3     & 21$\pm$2    & 6$\pm$4:    \\ 
%
\enddata
\tablecomments{
See Table~\ref{tab_spec_bel} for details.
}
\end{deluxetable}

\begin{deluxetable}{ccrlllllr} 
\tablecaption{Spectroscopic Data for CNOC2 Absorption-Line AGN Candidates\label{tab_spec_abs}}
\rotate
\tabletypesize{\small}
\tablewidth{575.00000pt}
\tablenum{3}
\tablehead{
\colhead{}   & \colhead{}    & \colhead{}    & \colhead{\feii} & \colhead{\feii} & \colhead{\feii} & \colhead{\feii} & \colhead{\mgii} & \colhead{}      \\[.2ex]
\colhead{ID} & \colhead{$z$} & \colhead{$R_{cor}$} & \colhead{2255}           & \colhead{2344} & \colhead{2380} & \colhead{2600}  & \colhead{2798}  & \colhead{Other}
}
\startdata
0920.110823 & 0.73337$\pm$0.00046 & 3.75 & \nodata        & \nodata         & \nodata         & $-$5.6$\pm$5.2  & $-$11.6$\pm$6.0 & \nodata \\
2148.162196 & 0.74842$\pm$0.00014 & 5.00 & \nodata        & \nodata         & \nodata         & $-$8.5$\pm$2.6  & $-$3.0$\pm$2.2  & \nodata \\
2148.150370 & 0.75621$\pm$0.00010 & 5.00 & \nodata        & \nodata         & \nodata         & $-$8.9$\pm$3.3: & $-$10.6$\pm$4.5 & \feii*\,$\lambda$2747, $-$3.3$\pm$2.0 \\
0920.060299 & 0.79999$\pm$0.00026 & 3.85 & \nodata        & \nodata         & \nodata         & $>-$7.8         & $-$13.1$\pm$4.0 & \nodata \\
0223.190667 & 0.84700$\pm$0.00010 & 5.00 & \nodata        & \nodata         & \nodata         & $-$6.7$\pm$5.0  & $-$7.6$\pm$4.2  & \nodata \\
2148.121113 & 0.88595$\pm$0.00039 & 3.11 & \nodata        & $>-$9.9:        & $>-$7.4         & $>-$5.8:        & $-$6.4$\pm$5.0  & \nodata \\
0920.101758 & 0.89979$\pm$0.00042 & 4.06 & \nodata        & $>-$4.9         & $>-$6.8         & $-$4.8$\pm$4.9  & $-$2.9$\pm$4.2  & \nodata \\
2148.130511 & 0.91352$\pm$0.00043 & 3.88 & \nodata        & $>-$6.7         & $>-$5.4         & $-$5.2$\pm$4.0  & $-$10.0$\pm$3.1 & \nodata \\
1447.140898 & 1.01769$\pm$0.00017 & 5.00 & $>-$2.4        & $>-$2.2         & $>-$2.3         & $-$5.8$\pm$2.1  & $-$4.2$\pm$2.0  & \nodata \\
0223.071845 & 1.06096$\pm$0.00010 & 5.00 & $>-$6.3        & $>-$5.2         & $>-$5.2         & $-$6.4$\pm$4.3  & $-$11.0$\pm$3.7 & \nodata \\
1447.150061 & 1.07162$\pm$0.00037 & 4.88 & $>-$3.3        & $>-$1.8         & $-$1.4$\pm$1.8  & $-$4.5$\pm$3.2: & $-$5.6$\pm$3.5: & \nodata \\
1447.020849 & 1.07458$\pm$0.00041 & 3.93 & $-$8.5$\pm$4.7 & $-$3.5$\pm$4.2  & $-$5.7$\pm$4.1  & $-$8.1$\pm$3.7  & $-$12.9$\pm$8.7 & \nodata \\
1447.091045 & 1.11503$\pm$0.00032 & 3.62 & $>-$5.2        & $-$2.9$\pm$3.8  & $-$3.8$\pm$3.8: & $-$10.6$\pm$3.9 & \nodata         & \nodata \\
1447.111559 & 1.17736$\pm$0.00044 & 4.53 & $>-$3.7        & $-$1.9$\pm$1.9: & $-$2.2$\pm$2.5: & $-$4.9$\pm$4.4  & $-$6.7$\pm$3.6  & \nodata \\
2148.050488 & 1.32701$\pm$0.00013 & 5.00 & $>-$2.25       & $-$2.3$\pm$1.6: & $-$4.7$\pm$2.0: & $-$7.7$\pm$2.1  & \nodata         & \CIII?, 30$\pm$7: \\ 
            & \nodata          & \nodata &          &          &          &         &         & H$\gamma$, 27$\pm$15: \\ 
            & \nodata          & \nodata &          &          &          &         &         & H$\beta$, 150$\pm$40 \\ 
            & 1.32800$\pm$0.00100 & 3.59 &          &          &          &         &         & \OIII\,$\lambda$4959, 9$\pm$2: \\ 
            & 1.32800$\pm$0.00100 &      &          &          &          &         &         & \OIII\,$\lambda$5007, 40$\pm$14 \\ 
            & 1.32365$\pm$0.00060 & \nodata &          &          &          &         &         & H$\alpha$, 340$\pm$60 \\ 
\enddata
\tablecomments{
See Table~\ref{tab_spec_bel} for details.
\feii\,2255 denotes lines at 2249.88 \& 2260.78\,\AA,
\feii\,2380 lines at 2374.46 \& 2382.77\,\AA\ and
\feii\,2600 lines at 2586.65 \& 2600.17\,\AA\ 
(possibly confused with Mn\,{\sc ii} at 2576.88, 2594.50 and 2606.46\,\AA).
For equivalent width measurements, upper limits are denoted by $>$ for
absorption lines (negative EWs) and $<$ for emission lines (positive EWs).
The $R_{cor}$ entries for objects used to construct the high-$z$
absorption-line template have been set to 5.00.  
The redshift uncertainties for these objects will be underestimates.
For 2148.050488, several emission as well as absorption lines are listed:
\CIII\ may be present in the CNOC2 discovery spectrum, and the remaining
lines are from an infrared spectrum.  
Redshifts are given for the \OIII\ lines (the adopted systemic redshift)
and the blueshifted H$\alpha$ line; 
the latter has no $R_{cor}$ value due to the lack of a suitable template.
}
\end{deluxetable}

\begin{deluxetable}{ccccrcccccccccc} 
\tablecaption{Photometric Data for CNOC2 AGN\label{tab_phot}}
\rotate
\tabletypesize{\small}
\tablewidth{635.00000pt}
\tablenum{4}
\tablehead{
\colhead{ID} & \colhead{IAU Designation} & \colhead{z} & \colhead{Pcl} & \colhead{$M_B$}& \colhead{U} & \colhead{err} & \colhead{B} & \colhead{err} & \colhead{V} & \colhead{err} & \colhead{R} & \colhead{err} & \colhead{I} & \colhead{err}
}
\startdata
\multicolumn{15}{c}{Broad Emission-Line Objects}  \\
\hline
0223.190359 & J022355.5$-$000849 &  0.26970 & 1 & --19.87 & 20.30 & 0.04 & 20.85 & 0.04 & 20.20 & 0.04 & 19.53 & 0.03 & 18.90 & 0.03 \\
0223.130495\tablenotemark{a} & J022508.0$+$001707 &  0.52811 & 1 & --22.99 & 18.52 & 0.03 & 18.93 & 0.03 & 18.59 & 0.03 & 18.27 & 0.03 & 17.70 & 0.03 \\ 
2148.151660 & J215033.6$-$054733 &  0.60233 & 1 & --21.08 & 20.95 & 0.19 & 21.08 & 0.04 & 20.92 & 0.04 & 20.36 & 0.04 & 19.58 & 0.04 \\
0920.081010 & J092425.4$+$370659 &  0.64744 & 3 & --21.85 & 19.98 & 0.04 & 20.45 & 0.03 & 20.08 & 0.03 & 19.75 & 0.03 & 19.09 & 0.04 \\
2148.102026\tablenotemark{ad} & J215200.4$-$054524 &  0.65243 & 3\tablenotemark{b} & --21.02 & 21.16 & 0.08 & 21.30 & 0.04 & 20.93 & 0.04 & 20.31 & 0.04 & 19.30 & 0.04 \\
1447.170990\tablenotemark{d} & J144839.9$+$084253 &  0.66707 & 3 & --20.45 & 21.78 & 0.10 & 21.91 & 0.05 & 21.49 & 0.05 & 20.98 & 0.04 & 20.10 & 0.05 \\ 
1447.051040 & J144948.8$+$093836 &  0.69900 & 2 & --20.12 & 21.68 & 0.12 & 22.34 & 0.07 & 22.13 & 0.07 & 21.53 & 0.05 & 21.01 & 0.07 \\ 
1447.050689 & J144926.9$+$093710 &  0.70100 & 2 & --20.04 & 21.58 & 0.26 & 22.41 & 0.07 & 22.19 & 0.07 & 22.07 & 0.06 & 21.52 & 0.09 \\ 
2148.040045 & J215103.3$-$051408 &  0.72774 & 1 & --22.49 & 19.56 & 0.04 & 20.04 & 0.03 & 19.55 & 0.03 & 19.38 & 0.03 & 18.76 & 0.03 \\
2148.031234 & J215112.9$-$051843 &  0.73200 & 1 & --21.62 & 20.38 & 0.05 & 20.92 & 0.04 & 20.86 & 0.04 & 20.31 & 0.04 & 19.65 & 0.04 \\
0920.182673 & J092148.6$+$364248 &  0.74002 & 1 & --21.05 & 21.11 & 0.11 & 21.51 & 0.04 & 20.91 & 0.04 & 20.00 & 0.04 & 19.44 & 0.04 \\
1447.161148 & J144825.1$+$085249 &  0.77000 & 2\tablenotemark{b} & --20.86 & 21.29 & 0.08 & 21.79 & 0.04 & 21.45 & 0.04 & 21.20 & 0.05 & 20.62 & 0.05 \\ 
1447.090674 & J144958.6$+$085912 &  0.79000 & 2 & --21.06 & 21.37 & 0.08 & 21.63 & 0.04 & 21.11 & 0.04 & 20.77 & 0.04 & 19.92 & 0.05 \\ 
1447.081984 & J145011.0$+$091115 &  0.84100 & 3 & --20.71 & 21.39 & 0.12 & 22.10 & 0.05 & 21.84 & 0.04 & 21.61 & 0.06 & 20.91 & 0.05 \\
0920.080251\tablenotemark{c} & J092445.4$+$370245 &  0.86100 & 1\tablenotemark{b} & --20.32 & 22.05 & 0.10 & 22.51 & 0.05 & 21.90 & 0.04 & 21.83 & 0.07 & 21.10 & 0.11 \\ 
0223.021739 & J022557.6$+$002912 &  0.88600 & 3 & --21.90 & 20.93 & 0.06 & 21.02 & 0.04 & 20.81 & 0.04 & 20.61 & 0.04 & 20.20 & 0.04 \\
2148.110259 & J215130.6$-$055251 &  0.88800 & 2 & --21.05 & 21.49 & 0.09 & 21.87 & 0.05 & 21.93 & 0.05 & 21.56 & 0.04 & 21.56 & 0.10 \\
0920.140041 & J092309.0$+$365254 &  0.94000 & 2 & --20.53 & 21.75 & 0.12 & 22.50 & 0.07 & 21.90 & 0.05 & 21.68 & 0.05 & 21.06 & 0.06 \\
2148.011928 & J215118.2$-$052956 &  1.06300 & 1 & --20.43 & 21.97 & 0.09 & 22.79 & 0.07 & 21.39 & 0.05 & 20.99 & 0.04 & 20.47 & 0.08 \\
1447.090792 & J145016.2$+$085938 &  1.12000 & 3 & --22.48 & 20.25 & 0.04 & 20.82 & 0.04 & 20.37 & 0.03 & 19.97 & 0.03 & 19.62 & 0.05 \\ 
1447.150051 & J144900.1$+$084832 &  1.16508 & 2 & --21.46 & 21.30 & 0.08 & 21.90 & 0.06 & 21.31 & 0.04 & 20.90 & 0.04 & 20.56 & 0.09 \\
2148.060762 & J215100.6$-$045305 &  1.38000 & 1 & --22.29 & 20.16 & 0.05 & 21.33 & 0.08 & 20.42 & 0.09 & 20.15 & 0.10 & 18.55\tablenotemark{e} & 0.13 \\
1447.162128 & J144819.0$+$085629 &  1.77000 & 2\tablenotemark{b} & --22.04 & 21.69 & 0.09 & 21.94 & 0.07 & 21.95 & 0.06 & 21.61 & 0.04 & 21.25 & 0.08 \\
0223.151505 & J022514.5$+$000544 &  1.82835 & 2\tablenotemark{b} & --22.59 & 20.90 & 0.06 & 21.45 & 0.04 & 21.05 & 0.04 & 20.93 & 0.04 & 20.44 & 0.07 \\
0920.020246 & J092407.1$+$370947 &  1.94300 & 3 & --23.58 & 20.33 & 0.04 & 20.56 & 0.04 & 20.43 & 0.04 & 20.16 & 0.03 & 19.93 & 0.04 \\
1447.131920 & J144857.2$+$091230 &  1.96500 & 2 & --21.90 & 21.54 & 0.10 & 22.31 & 0.06 & 22.13 & 0.09 & 22.09 & 0.08 & 21.31 & 0.06 \\ 
1447.171440\tablenotemark{c} & J144824.9$+$084419 &  2.00000 & 2\tablenotemark{b} & --23.13 & 21.89 & 0.12 & 21.06 & 0.06 & 21.17 & 0.04 & 21.24 & 0.02 & 21.09 & 0.02 \\ 
1447.132036 & J144907.7$+$091300 &  2.02000 & 2 & --21.76 & 21.85 & 0.12 & 22.41 & 0.09 & 22.33 & 0.12 & 22.09 & 0.09 & 21.38 & 0.08 \\ 
0223.030993 & J022547.9$+$003330 &  2.09200 & 3 & --22.67 & 20.76 & 0.05 & 21.58 & 0.04 & 21.34 & 0.04 & 21.10 & 0.04 & 20.75 & 0.08 \\
0920.092046 & J092420.5$+$370119 &  2.45000 & 3 & --22.67 & 22.18 & 0.11 & 21.78 & 0.05 & 21.47 & 0.04 & 21.32 & 0.04 & \nodata & \nodata \\
0223.051932 & J022545.7$+$005147 &  2.51571 & 3 & --23.09 & 21.41 & 0.09 & 21.43 & 0.04 & 21.10 & 0.07 & 20.87 & 0.04 & 20.60 & 0.09 \\
\\ \\ \\ \\ \\ \\
1447.051899\tablenotemark{d} & J144932.8$+$094207 &  3.08000 & 2 & --22.12 & 23.15 & 0.36 & 23.05 & 0.09 & 21.75 & 0.06 & 20.86 & 0.04 & 20.02 & 0.04 \\
2148.112371 & J215136.6$-$054552 &  3.15700 & 3 & --22.31 & 22.94 & 0.28 & 22.97 & 0.12 & 21.82 & 0.08 & 21.35 & 0.05 & 20.97 & 0.11 \\ 
1447.020157\tablenotemark{d} & J144925.2$+$091245 &  3.17000 & 3 & --22.85 & 22.36 & 0.50 & 22.45 & 0.08 & 21.88 & 0.05 & 21.43 & 0.04 & 20.74 & 0.05 \\ 
2148.171412 & J215018.2$-$055550 &  3.19543 & 1 & --22.21 & 22.19 & 0.50 & 23.13 & 0.14 & 21.99 & 0.09 & 21.74\tablenotemark{f} & 0.07 & 21.52\tablenotemark{f} & 0.27 \\ 
0223.151510\tablenotemark{c} & J022532.3$+$000548 &  3.44990 & 1\tablenotemark{b} & --21.59 & 24.04 & 0.39 & 24.18 & 0.24 & 22.90 & 0.08 & 22.71 & 0.12 & 22.02 & 0.13 \\ 
0223.010767 & J022606.5$+$001742 &  4.00188 & 2\tablenotemark{b} & --23.11 & 22.41 & 0.50 & 24.16 & 0.15 & 22.06 & 0.06 & 21.10 & 0.04 & 20.52 & 0.05 \\ 
0223.011745 & J022612.0$+$002156 &  4.67550 & 3 & --23.56 & 23.53 & 0.33 & 25.31 & 0.41 & 22.32 & 0.09 & 20.72 & 0.04 & 20.22 & 0.04 \\
\hline
\multicolumn{15}{c}{[Ne\,{\sc v}] Objects} \\
\hline
2148.170418\tablenotemark{d} & J214949.5$-$060031 & 0.31874 & 1 & --19.52 & 21.90 & 0.13 & 21.98 & 0.08 & 20.53 & 0.04 & 19.66 & 0.04 & 19.14 & 0.04 \\ 
0223.040997\tablenotemark{d} & J022617.3$+$004114 & 0.39868 & 1 & --18.85 & 22.44 & 0.12 & 22.56 & 0.06 & 21.37 & 0.05 & 20.43 & 0.03 & 19.51 & 0.04 \\
1447.150676\tablenotemark{d} & J144910.1$+$085214 & 0.46667 & 1 & --20.34 & 21.41 & 0.09 & 21.35 & 0.04 & 20.47 & 0.04 & 19.74 & 0.04 & 19.11 & 0.04 \\
1447.080513\tablenotemark{d} & J145012.8$+$090620 & 0.48287 & 2 & --18.98 & 22.96 & 0.36 & 23.44 & 0.17 & 22.06 & 0.06 & 20.88 & 0.04 & 20.08 & 0.04 \\ 
1447.140343\tablenotemark{d} & J144911.5$+$085829 & 0.50796 & 1 & --19.70 & 23.76 & 0.50 & 22.83 & 0.11 & 21.44 & 0.06 & 20.38 & 0.04 & 19.43 & 0.04 \\ 
0223.130940 & J022509.6$+$001904 &  0.53942 & 3 & --20.91 & 20.62 & 0.05 & 21.05 & 0.04 & 20.70 & 0.03 & 20.26 & 0.03 & 19.38 & 0.04 \\
0920.161307\tablenotemark{d} & J092237.9$+$364827 & 0.56147 & 1 & --18.95 & 22.29 & 0.19 & 23.09 & 0.09 & 22.00 & 0.06 & 21.04 & 0.06 & 20.01 & 0.04 \\
0920.020116\tablenotemark{d} & J092356.8$+$370914 &  0.56225 & 1 & --20.24 & 22.28 & 0.10 & 21.80 & 0.05 & 21.27 & 0.04 & 20.54 & 0.04 & 19.66 & 0.04 \\
\hline
\\ \\ \\ \\ \\ \\ \\ \\ \\ \\ \\ \\ \\ \\ \\ \\ \\ \\ \\ \\ \\ \\ 
\multicolumn{15}{c}{Absorption-Line Objects} \\
\hline
0920.110823 & J092346.6$+$364755 &  0.73337 & 1 & --20.41 & 21.68 & 0.09 & 22.13 & 0.04 & 21.62 & 0.05 & 21.07 & 0.05 & 20.40 & 0.04 \\
2148.162196 & J214959.0$-$054603 &  0.74842 & 1 & --21.01 & 21.17 & 0.10 & 21.57 & 0.05 & 21.30 & 0.05 & 20.82 & 0.04 & 20.26 & 0.05 \\
2148.150370 & J215045.5$-$055227 &  0.75621 & 1 & --21.39 & 20.60 & 0.07 & 21.22 & 0.04 & 20.75 & 0.04 & 20.33 & 0.04 & 19.64 & 0.07 \\
0920.060299 & J092409.1$+$373918 &  0.79999 & 1 & --20.45 & 21.57 & 0.08 & 22.27 & 0.05 & \nodata & \nodata & 20.97 & 0.05 & 20.14 & 0.05 \\
0223.190667 & J022349.5$-$000754 &  0.84700 & 1 & --21.29 & 20.93 & 0.07 & 21.54 & 0.04 & 21.09 & 0.05 & 20.58 & 0.04 & 19.88 & 0.04 \\
2148.121113 & J215124.9$-$054013 &  0.88595 & 1 & --20.80 & 21.51 & 0.11 & 22.12 & 0.05 & 22.00 & 0.07 & 21.80 & 0.08 & 21.03 & 0.10 \\
0920.101758 & J092446.1$+$365233 &  0.89979 & 1 & --20.57 & 22.08 & 0.16 & 22.38 & 0.09 & 21.54 & 0.09 & 21.58 & 0.09 & 20.26 & 0.09 \\
2148.130511 & J215025.0$-$053428 &  0.91352 & 1 & --21.78 & 20.75 & 0.07 & 21.19 & 0.05 & 20.89 & 0.04 & 20.57 & 0.04 & 19.93 & 0.04 \\
1447.140898 & J144850.5$+$090216 &  1.01769 & 2 & --21.43 & 20.95 & 0.06 & 21.73 & 0.05 & 21.36 & 0.04 & 21.21 & 0.07 & 20.61 & 0.07 \\
0223.071845 & J022616.0$+$010759 &  1.06096 & 2 & --20.85 & \nodata & \nodata & 22.38 & 0.06 & 22.12 & 0.13 & 21.67 & 0.05 & 20.86 & 0.08 \\
1447.150061 & J144850.8$+$084836 &  1.07162 & 3 & --21.13 & 21.47 & 0.09 & 22.11 & 0.06 & 21.88 & 0.05 & 21.71 & 0.07 & 21.28 & 0.10 \\
1447.020849 & J144945.9$+$091650 &  1.07458 & 2 & --21.00 & 21.57 & 0.11 & 22.24 & 0.06 & 21.99 & 0.05 & 21.67 & 0.05 & 20.98 & 0.07 \\
1447.091045 & J145029.8$+$090045 &  1.11503 & 1 & --21.43 & 21.06 & 0.07 & 21.86 & 0.04 & 21.45 & 0.05 & 21.16 & 0.04 & 20.60 & 0.06 \\
1447.111559 & J144930.0$+$085541 &  1.17736 & 1 & --21.14 & 21.41 & 0.09 & 22.24 & 0.09 & 21.95 & 0.05 & 21.64 & 0.07 & 21.04 & 0.08 \\
2148.050488 & J215112.5$-$050414 &  1.32800 & 3 & --22.30 & 20.64 & 0.07 & 21.26 & 0.04 & 20.65 & 0.04 & 20.23 & 0.04 & 19.83 & 0.04 \\ 
\enddata
\tablenotetext{a}{Object is a radio source.  
For details see the notes on each object in the text.}
\tablenotetext{b}{Classification may be in error, with the actual appropriate
Pcl value being 3 (or 2 for 2148.102026).}  
\tablenotetext{c}{Object's magnitudes were contaminated by a very close neighbor
galaxy not separated from the AGN by PPP in the CNOC2 pipeline.  The magnitudes
have been corrected in this table for those neighbors, but remain more
uncertain than those of other objects due to the small apertures used,
from which large corrections are needed to reach total magnitudes.}
\tablenotetext{d}{Object would not qualify as a color-selected AGN candidate
using criteria based on those of \citet{hal96dms2}.}  
\tablenotetext{e}{Object's $I$ magnitude is probably affected by scattered
light from a nearby bright star.}
\tablenotetext{f}{Object's $R$ and $I$ magnitudes may be affected by a bleeding
column from a nearby bright star.}
\tablecomments{The IAU Designation for each object gives the RA and
Dec coordinates in epoch J2000, and should be preceded by the prefix ``CNOC2".
The Pcl parameter is the morphological object class assigned by PPP:
galaxies are class 1, probable galaxies class 2, stars class 3, 
and saturated stars class 4.
A magnitude error of 0.50 indicates 
a $2\sigma$ upper limit.}
\end{deluxetable}

\clearpage

\end{document}